\documentclass[lettersize,journal]{IEEEtran}
\IEEEoverridecommandlockouts
\usepackage{makecell}
\usepackage{array}
\usepackage{graphicx,amssymb,amsmath}
\usepackage{multicol}
\usepackage[noadjust]{cite}
\usepackage{setspace}
\usepackage{subfigure}
\usepackage{graphicx}
\usepackage{float}
\usepackage{url}
\usepackage{stfloats}
\usepackage{amsthm,pifont}
\usepackage{flushend}
\usepackage{cases,subeqnarray}
\usepackage{bm,multirow,bigstrut}
\usepackage{amsmath, amsthm, amssymb}
\usepackage{textcomp}
\usepackage{latexsym,bm}
\usepackage{booktabs}
\usepackage{xcolor}
\usepackage{mathtools}
\usepackage{dsfont}
\usepackage{extarrows}
\usepackage{epsfig}
\usepackage{epstopdf}
\usepackage[noend]{algpseudocode}
\usepackage{algorithmicx,algorithm}

\usepackage{algorithm}
\usepackage{algpseudocode}
\usepackage{amsmath}

\usepackage{cite}
\usepackage{bm}
\usepackage{cleveref}
\usepackage{multicol}       % table
\usepackage{multirow}       % table
\usepackage{array}          % table
\usepackage{colortbl}
\usepackage{makecell}
\definecolor{crimson}{RGB}{192,0,0}         % color crimson
\definecolor{navy}{RGB}{47,85,151}         % color crimson
\usepackage{bbding}
\usepackage{graphicx}
\usepackage{booktabs}
\usepackage{algorithm}
\usepackage{algpseudocode}

\theoremstyle{plain}

\theoremstyle{plain}
\newtheorem{rem}{Remark}
\newtheorem{them}{Theorem}

\usepackage{amsmath}

\IEEEoverridecommandlockouts
\begin{document}

%----------------------------title&author&tahnks----------------------------
%\title{Stacked Intelligent Metasurfaces to Reduce Power Consumption and Cost for Cell-Free Massive MIMO Systems}

\title{Harnessing Stacked Intelligent Metasurface for Enhanced Cell-Free Massive MIMO Systems: A Low-Power and Cost Approach}
%\title{Uplink Performance of Stacked Intelligent Metasurface-enhanced Cell-Free Massive MIMO Systems}
%\title{Enhanced Cell-Free Massive MIMO Systems With Stacked Intelligent Metasurfaces}
\author{Enyu Shi,~\IEEEmembership{Graduate Student Member,~IEEE}, Jiayi~Zhang,~\IEEEmembership{Senior Member,~IEEE}, Yiyang Zhu,~\IEEEmembership{Student Member,~IEEE} Jiancheng~An,~\IEEEmembership{Member,~IEEE}, Chau Yuen,~\IEEEmembership{Fellow,~IEEE}, and Bo Ai,~\IEEEmembership{Fellow,~IEEE}
%\IEEEauthorblockA{School of Electronic and Information Engineering, Beijing Jiaotong University, Beijing 100044}
%\thanks{This work was supported in part by National Key R\&D Program of China under Grant 2020YFB1806903, in part by National Natural Science Foundation of China under Grants 61971027, in part by Beijing Natural Science Foundation under Grant L202013, in part by Frontiers Science Center for Smart High-speed Railway System. D. W. K. Ng is supported by the Australian Research Council's Discovery Project (DP210102169, DP230100603). (\emph{Corresponding author: Jiayi Zhang.})}
\thanks{E. Shi, J. Zhang, Y. Zhu, and B. Ai are with the School of Electronics and Information Engineering, Beijing Jiaotong University, Beijing 100044, P. R. China. (e-mail: \{enyushi, jiayizhang, yiyangzhu, boai\}@bjtu.edu.cn).}
\thanks{J. An and C. Yuen are with the School of Electrical and Electronics Engineering, Nanyang Technological University, Singapore 639798 (e-mail: jiancheng\_an@163.com, chau.yuen@ntu.edu.sg).}
%\thanks{D. W. K. Ng is with the School of Electrical Engineering and Telecommunications, University of New South Wales, NSW 2052, Australia. (e-mail: w.k.ng@unsw.edu.au).}
%\thanks{B. Ai is with the State Key Laboratory of Rail Traffic Control and Safety, Beijing Jiaotong University, Beijing 100044, China. (e-mail: boai@bjtu.edu.cn).}
}

\maketitle
%\vspace{-17mm}
%\thispagestyle{empty}
%\pagestyle{empty}
%----------------------------abstract----------------------------
\begin{abstract}
In this paper, we explore the integration of low-power, low-cost stacked intelligent metasurfaces (SIM) into cell-free (CF) massive multiple-input multiple-output (mMIMO) systems to enhance access point (AP) capabilities and address high power consumption and cost challenges.
Specifically, we investigate the uplink performance of a SIM-enhanced CF mMIMO system and propose a novel system framework. 
First, the closed-form expressions of the spectral efficiency (SE) are obtained using the unique two-layer signal processing framework of CF mMIMO systems. Second, to mitigate inter-user interference, an interference-based greedy algorithm for pilot allocation is introduced. Third, a wave-based beamforming algorithm for SIM is proposed, based only on statistical channel state information, which effectively reduces the fronthaul costs. Finally, a max-min SE power control algorithm is proposed to improve the performance of UE with inferior channel conditions.
The results indicate that increasing the number of SIM layers and meta-atoms leads to significant performance improvements and allows for a reduction in the number of APs and AP antennas, thus lowering the costs.
In particular, the best SE performance is achieved with the deployment of 20 APs plus 1200 SIM meta-atoms.
Finally, the proposed wave-based beamforming algorithm can enhance the SE performance of SIM-enhanced CF-mMIMO systems by 57\%, significantly outperforming traditional CF mMIMO systems. 
\end{abstract}

\begin{IEEEkeywords}
Stacked intelligent metasurface, cell-free massive MIMO, wave-based beamforming, power control, spectral efficiency.
\end{IEEEkeywords}

\IEEEpeerreviewmaketitle

%\vspace{-0.2cm}
\section{Introduction}
\IEEEPARstart{T}{he} imminent sixth-generation (6G) network is expected to be crucial in various sectors of future society, industry, and daily life, demanding stringent requirements for communication in terms of capacity, latency, reliability, and intelligence \cite{wang2023road}. The cell-free (CF) massive multiple-input multiple-output (mMIMO) system, presented as a viable replacement for cellular networks, exhibits significant advantages in ultra-dense networks due to its distributed network architecture. In CF-mMIMO systems, massive access points (APs) linked to the central processing unit (CPU) via fronthaul are uniformly distributed across the service area, achieving spatial diversity while reducing the distance between transmitters and receivers \cite{ngo2017cell}. The APs collaborate to simultaneously provide communication services to user equipment (UEs), leading to an improvement in the quality of service (QoS).

Although the introduction of CF mMIMO improves system performance, the deployment of numerous APs has escalated concerns about energy consumption and hardware costs in CF mMIMO systems. To address these challenges, reconfigurable intelligent surfaces (RISs) are introduced as a promising solution capable of improving the performance of CF mMIMO systems while reducing energy and costs \cite{9743355,ma2022cooperative,wu2019intelligent}. Many studies have investigated how to integrate CF and RIS to fully leverage the advantages of both.
For example, the authors in \cite{10167480} studied the uplink spectral efficiency (SE) of the RIS-assisted CF mMIMO system with electromagnetic interference (EMI) and proposed a fractional power control method to resist communication performance degradation. Besides, \cite{zhang2021joint} proposed a joint RIS phase shift and AP precoding framework of a wideband RIS-assisted CF network to maximize the sum rate. The results demonstrate that significant improvements in system throughput can be achieved through RIS phase shift design, and positioning RIS closer to APs can yield better performance. Also, the authors in \cite{le2021energy} introduced RIS to reduce the energy consumption of CF mMIMO systems and to maximize system energy efficiency (EE) through the optimization of RIS phase shifts.
Extensive studies have validated that the strategic design of RIS phase shift can enhance system performance and save energy consumption. Nonetheless, the inherently dynamic characteristics of wireless environments necessitate frequent execution of the joint AP-RIS beamforming optimization, typically resulting in escalated signal processing complexity \cite{an2021low,10129196}. Moreover, the quasi-passive nature of single-layer RIS, compounded by hardware constraints and significant path loss, restricts its capacity for advanced MIMO functionality implementation. 

Fortunately, motivated by the rapid development of metasurface design, recent research on intelligent surfaces has proposed to adopt multi-layer metasurfaces for signal processing within the electromagnetic (EM) wave domain \cite{doi:10.1126/science.aat8084,liu2022programmable,nerini2024physically}. For example, in \cite{liu2022programmable}, the authors developed a programmable diffractive deep neural network structure using a multilayer metasurface array, in which each meta-atom functions as a reconfigurable artificial neuron. Inspired by this, the authors in \cite{10158690} introduced a novel stacked intelligent metasurface (SIM)-assisted MIMO transceiver. This approach layers multiple nearly passive, programmable metasurfaces to create a SIM with a configuration akin to that of an artificial neural network (ANN), thus significantly enhancing signal processing capabilities. Moreover, the authors in \cite{nadeem2023hybrid,papazafeiropoulos2024performance,wang2024multi} exploited SIM to enable receiver combining and transmit precoding in holographic MIMO communications. The results indicate that compared to a single-layer RIS, SIM can achieve better system performance. Furthermore, the authors in \cite{yao2024channel,an2024two} investigate the channel and direction-of-arrival estimation to promote the development of SIM.
Consequently, SIMs enable the deployment of MIMO transceivers that feature sophisticated wave domain transmit precoding and receive combining, while markedly decreasing RF energy usage and hardware expenses \cite{an2023stacked3,liu2024drl}. 

Hence, substituting APs with SIM can leverage the advantages of both, addressing the issues of high power consumption by multiple APs and active antennas in CF mMIMO systems, thereby enhancing system performance. However, the current studies are predicated on the assumption of known perfect instantaneous channel state information (CSI). For example, in \cite{an2023stacked,hassan2024efficient}, the author utilized instantaneous channel state information (CSI) to optimize and update the wave-based beamforming of SIM for maximizing the sum rate. 
Given the large number of meta-atoms in SIM, phase shift design based on instantaneous CSI necessitates the SIM controller having ultra-high-speed computation capability, presenting a significant challenge for CF mMIMO systems. Especially in CF mMIMO systems, the presence of multiple APs and enhanced SIM leads to greater computational complexity and poses higher demands on the fronthaul links in terms of capacity and data rate \cite{masoumi2019performance}. Therefore, devising an analytical framework based on statistical CSI to explore the performance of SIM integrated with CF mMIMO systems would be immensely beneficial for the practical implementation of such systems.

Motivated by the observation above, we propose the concept of SIM-enhanced CF mMIMO systems to improve performance with low cost and power consumption. The central idea involves substituting the conventional AP with a SIM-enhanced AP. Then, we investigate the considered system performance by adopting the two-stage signal processing architecture of the SIM-enhanced CF mMIMO system and obtain the closed-form expression for uplink SE with statistical CSI. Then, we designed algorithms based on the SE closed-form expression for wave-based beamforming optimization of SIM and power control to unlock the system potential. To our best knowledge, this constitutes the first endeavor to implement SIM within CF mMIMO networks. The contributions of this work are delineated as follows:

\begin{itemize}
\item We propose the concept of integrating low-power and low-cost SIM into the CF mMIMO, namely SIM-enhanced CF mMIMO, to further enhance the system performance. The key idea is to integrate SIMs in front of each AP to replace some of the AP antennas required in CF mMIMO systems. We investigate the uplink SE performance of the SIM-enhanced CF mMIMO system with spatially correlated channels and statistical CSI. The MMSE estimator is adopted for channel estimation. In practice, we propose the two-stage signal processing, i.e., the maximum ratio (MR) combining at the APs and the centralized large-scale fading decoding (LSFD)/equal gain combining decoding (EGCD) at the CPU. Then, the \textbf{closed-form expressions} for uplink SE are obtained for both the LSFD and EGCD.

\item To unlock the proposed system potential, we propose some algorithms for system optimization design. First, to mitigate inter-user interference, we propose an interference-based greedy algorithm for \textbf{pilot allocation}. Second, to unlock the potential of SIM, we propose an iterative optimization algorithm for designing the \textbf{wave-based beamforming} to configure SIMs only once in each coherence block, which effectively reduces the fronthaul costs based on the derived SE closed-form expressions. Furthermore, we propose a max-min SE \textbf{power control} algorithm to improve the performance of UEs with inferior channel conditions and improve fairness.

\item Our analysis explores the impact of various SIM and CF mMIMO system parameters on performance outcomes, including the number of SIM metasurface layers, meta-atoms, AP antennas, and different algorithms. 
We find that increasing the number of SIM meta-atoms per layer is always beneficial. 
Interestingly, the performance gain caused by increasing the number of layers relies on the metasurface aperture. Specifically, when the number of meta-atoms per layer is less than 9, the increase in the number of SIM metasurface layers results in a decrease in SE. 
%Additionally, SIMs featuring a number of meta-atoms more than 36 can ensure the stability of system performance even in the absence of phase shift design. This suggests that high-density SIM meta-atoms deployments can effectively mitigate phase instability issues.
Also, for a 1200 total count of all SIM meta-atoms, there exists an optimal number of 20 APs for deployment. Beyond this optimal count, further increases in the number of APs do not necessarily enhance performance, indicating diminishing returns with excessive AP deployment.
Furthermore, the proposed wave-based beamforming algorithm can enhance the communication performance of SIM-enhanced CF-mMIMO systems by 57\%, significantly outperforming traditional CF systems. 
Additionally, the proposed power control algorithm improves the SE performance of 95\%-likely per-user by 70\% compared to full-power transmission.
SIMs, particularly those equipped with a large number of meta-atoms, offer a viable alternative to conventional AP antennas, enhancing the system robustness and reliability.
Our research offers theoretical guidance for the practical deployment of SIM in CF mMIMO systems.
\end{itemize}

The remainder of this paper is structured as follows. Section \uppercase\expandafter{\romannumeral2} outlines the SIM-enhanced CF mMIMO system model, considering spatial correlation and channel estimation errors, and details the uplink data transmission process. Next, Section \uppercase\expandafter{\romannumeral3} presents the SE closed-form expressions and performance analysis of the considered system with LSFD and EGCD. The SIM-enhanced CF mMIMO system configuration and design are proposed in \uppercase\expandafter{\romannumeral4}, including the pilot allocation, SIM wave-based beamforming, and power control. Then, numerical results and discussion are provided in Section \uppercase\expandafter{\romannumeral5}. Finally, Section \uppercase\expandafter{\romannumeral6} concludes this paper.

\textbf{Notation:} The column vectors and matrices are denoted by boldface lowercase letters $\mathbf{x}$ and boldface uppercase letters $\mathbf{X}$, respectively. The superscripts $\mathbf{x}^{\rm{H}}$, $x^\mathrm{T}$, and $x^\mathrm{*}$ are adopted to represent conjugate, transpose, and conjugate transpose, respectively. The $\triangleq$, $\left\|  \cdot  \right\|$, and $\left\lfloor  \cdot  \right\rfloor $ denote the definitions, the Euclidean norm, and the truncated argument, respectively. ${\rm{tr}}\left(  \cdot  \right)$, $\mathbb{E}\left\{  \cdot  \right\}$, and ${\rm{Cov}}\left\{  \cdot  \right\}$ denote the trace, expectation, and covariance operators, respectively. We exploit ${\text{diag}}\left( {{a_1}, \cdots ,{a_n}} \right)$ to express a diagonal matrix. Also, $\otimes$ and $\odot$ denote the Kronecker products and the element-wise products, respectively. 
The circularly symmetric complex Gaussian random variable $x$ with variance $\sigma^2$ is denoted by $x \sim \mathcal{C}\mathcal{N}\left( {0,{\sigma^2}} \right)$. Then, $\nabla$ denotes the gradient operation. $\mathbb{B}^n$, $\mathbb{Z}^n$, $\mathbb{R}^n$, and $\mathbb{C}^n$ represent the $n$-dimensional spaces of binary, integer, real, and complex numbers, respectively. Finally, the $N \times N$ zero matrix and identity matrix are denoted by $\mathbf{0}_{N}$ and $\mathbf{I}_{N}$, respectively.

%----------------------------system model----------------------------

\begin{figure*}[t]
\centering
\includegraphics[scale=0.9]{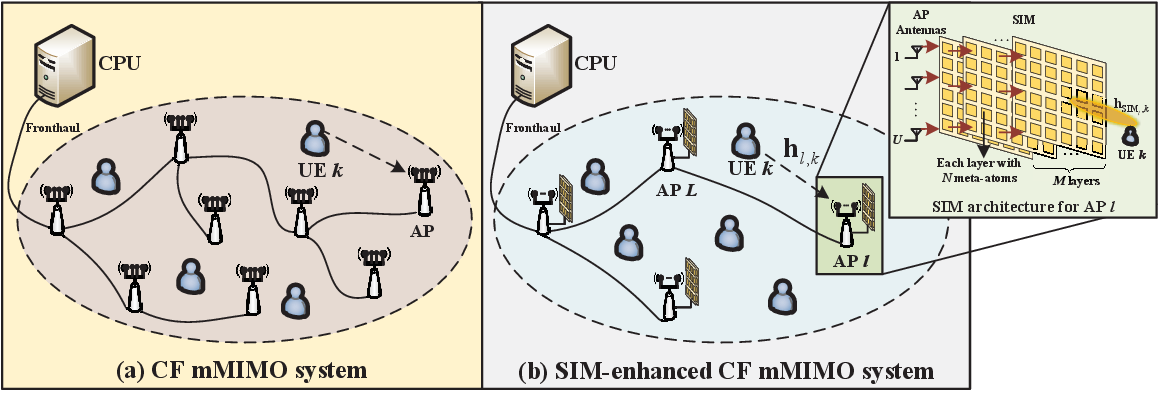}%\vspace{-0.2cm}
\caption{Illustration of CF mMIMO and SIM-enhanced CF mMIMO systems. (a) Illustration of the CF mMIMO system. (b) Illustration of the SIM-enhanced CF mMIMO system and the SIM architecture for AP $l$.}\label{system_model}\vspace{-0.3cm}
\end{figure*}

\vspace{-0.1cm}
\section{SIM-enhanced CF mMIMO System Model}\label{se:model}
As shown in Fig.~\ref{system_model} (b), we consider a SIM-enhanced CF mMIMO system that consists of $L$ SIM-enhanced APs, a central processing unit (CPU), and $K$ UEs.  We assume that each AP and UE is equipped with $U$ antennas and a single antenna, respectively. Specifically, we assume that each SIM is equipped with the same structure which has $M$ metasurface layers and $N$ meta-atoms in each layer. Let ${\cal L} = \{ 1, \ldots ,L\}$, ${\cal K} = \{ 1, \ldots ,K\}$, ${\cal U} = \{ 1, \ldots ,U\}$, ${\cal M} = \{ 1, \ldots ,M\}$, and ${\cal N} = \{ 1, \ldots ,N\}$ denote the index sets of APs, UEs, AP antennas, SIM metasurface layers, and meta-atoms per layer, respectively. Furthermore, the SIM is connected to an intelligent controller at the AP, capable of applying a distinct and tunable phase shift to the electromagnetic (EM) waves passing through each meta-atom \cite{an2023stacked2}. In this scenario, there is a one-to-one binding relationship between SIM and AP, which means that the number of SIMs is equal to the number of APs, denoted as $L$. All APs are connected to the CPU via fronthaul links which can send the AP data to the CPU for signal processing. In line with conventional CF mMIMO systems, it is postulated that the time division duplex (TDD) protocol is employed for the considered SIM-enhanced CF mMIMO system. 
Let $\tau_c$ represent the duration of each coherence block, where $\tau_p$ symbols are allocated for the uplink channel estimation phase, and $\tau_u = \tau_c - \tau_p$ symbols are employed for uplink data transmission phase.

\subsection{Channel Model}
We use ${e^{j\varphi _{l,m}^n}},\forall l \in {\cal L},\forall m \in {\cal M},\forall n \in {\cal N}$ with  $\varphi _{l,m}^n \in \left[ {0,2\pi } \right)$ denote the $n$-th meta-atom's phase shift in the $m$-th metasurface layer at the $l$-th SIM. Hence, the diagonal phase shift matrix ${{\bf{\Phi }}_{l,m}}$ for the $m$-th metasurface layer at the $l$-th SIM can be denoted as ${{\bf{\Phi }}_{l,m}} = {\rm{diag}}\left( {{e^{j\varphi _{l,m}^1}},{e^{j\varphi _{l,m}^2}}, \ldots ,{e^{j\varphi _{l,m}^N}}} \right) \in \mathbb{C} {^{N \times N}},\forall l \in {\cal L},m \in {\cal M}$. Furthermore, let ${\bf{W}}_{l,1}^{} = {\left[ {{\bf{w}}_{l,1}^1, \ldots ,{\bf{w}}_{l,1}^{U}} \right]^{\rm{T}}} \in \mathbb{C} {^{N \times U}}$ denote the transmission vector from the $l$-th AP to the first metasurface layer of the SIM, where ${{\bf{w}}_{l,1}^{{u}}} \in \mathbb{C} {^{N}}$ denotes the $u$-th antenna of AP $l$ to the first metasurface layer within the SIM. Let ${{\bf{W}}_{l,m}} \in \mathbb{C} {^{N \times N}},\forall m \ne 1,m \in {\cal M},l \in {\cal L}$ denote the transmission matrix from the $(m-1)$-th to the $m$-th metasurface layer of SIM $l$. According to the Rayleigh-Sommerfeld diffraction theory introduced by \cite{lin2018all,lin2024stacked}, the $\left( {n,n'} \right)$-th element is expressed as
%$w_{l,m}^{n,n'}$ of ${\bf{W}}_{l,1}^{}$

\begin{align}\label{w}
w_{l,m}^{n,n'} = \frac{{{d_x}{d_y}\cos \chi _{l,m}^{n,n'}}}{{d_{l,m}^{n,n'}}}\left( {\frac{1}{{2\pi d_{l,m}^{n,n'}}} - j\frac{1}{\lambda }} \right){e^{j2\pi \frac{{d_{l,m}^{n,n'}}}{\lambda }}},
\end{align}
where $\lambda$ represents the carrier wavelength, $d_{l,m}^{n,n'}$ indicates the corresponding transmission distance, $d_{x} \times d_{y}$ indicates the size of each SIM meta-atom, and ${\chi _{l,m}^{n,n'}}$ represents the angle between the propagation direction and the normal direction of the $(m-1)$-th SIM metasurface layer of SIM $l$. Similarly, the $n$-th element $w_{l,1,u}^u$ of ${{\bf{w}}_{l,1}^{{u}}}$ can be obtained from \eqref{w}.
Hence, the wave-based beamforming matrix ${\mathbf{G}_l} \in \mathbb{C}{^{N \times N}}$ of AP $l$, enabled by SIM, is obtained as 
\begin{align}\label{G_l}
{{\bf{G}}_l} = {{\bf{\Phi }}_{l,M}}{{\bf{W}}_{l,M}}{{\bf{\Phi }}_{l,M - 1}}{{\bf{W}}_{l,M - 1}} \ldots {{\bf{\Phi }}_{l,2}}{{\bf{W}}_{l,2}}{{\bf{\Phi }}_{l,1}}.
\end{align}

Then, we consider a quasi-static flat-fading channel model. 
Let ${\bf{h}}_{{\rm{SI}}{{\rm{M}}_l},k} \in \mathbb{C}{^{N \times 1}}$ denote the direct channel between the last metasurface layer of the SIM $l$ to UE $k$. Specifically, we assume that ${\bf{h}}_{{\rm{SI}}{{\rm{M}}_l},k}$ is characterized as the spatially correlated Rician fading channel, comprising a semi-deterministic line-of-sight (LoS) path component and a stochastic non-line-of-sight (NLoS) path component as
\begin{align}\label{h_simk}
{\bf{h}}_{{\rm{SI}}{{\rm{M}}_l},k}^{} = {{\bf{\Theta }}_{{\rm{SI}}{{\rm{M}}_l},k}}{\bf{\bar h}}_{{\rm{SI}}{{\rm{M}}_l},k}^{} + {\bf{g}}_{{\rm{SI}}{{\rm{M}}_l},k},
\end{align}
where ${\bf{\bar h}}_{{\rm{SI}}{{\rm{M}}_l},k}^{} \in \mathbb{C}{^{N \times 1}}$ denotes the deterministic LoS component and ${\bf{g}}_{{\rm{SI}}{{\rm{M}}_l},k}^{} \sim {\cal C}{\cal N}\left( {0,{{\bf{R}}_{{\rm{SI}}{{\rm{M}}_l},k}}} \right)$ denotes the NLoS component. ${{\bf{R}}_{{\rm{SI}}{{\rm{M}}_l},k}} = {\beta _{{\rm{SI}}{{\rm{M}}_l},k}}{\bf{R}} \in \mathbb{C}{^{N \times N}}$ where ${\beta _{{\rm{SI}}{{\rm{M}}_l},k}}$ denotes the distance-dependent path loss between UE $k$ and $l$-th SIM, and the covariance matrix ${\bf{R}} \in \mathbb{C}{^ {N \times N}}$ describes the spatial correlation among different meta-atoms of the output metasurface layer within the SIM. Without losing generality, we assume an isotropic scattering environment with multipath components uniformly distributed. Thus, the $\left( {n,n'} \right)$-th element of $\mathbf{R}$ is ${{\bf{R}}_{n,n'}} = {\rm{sinc}}\left( {2{{{d_{n,n'}}} \mathord{\left/
 {\vphantom {{{d_{n,n'}}} \lambda }} \right.
 \kern-\nulldelimiterspace} \lambda }} \right)$ \cite{bjornson2020rayleigh}, where $d_{n,n'}$ denotes the spacing distance between the meta-atoms and ${\rm{sinc}}\left( x \right) = {{\sin \left( {\pi x} \right)} \mathord{\left/
 {\vphantom {{\sin \left( {\pi x} \right)} {\left( {\pi x} \right)}}} \right.
 \kern-\nulldelimiterspace} {\left( {\pi x} \right)}}$ denotes the normalized sinc function.
Moreover, ${{\bf{\Theta }}_{{\rm{SI}}{{\rm{M}}_l},k}} = {\rm{diag}}\left( {{e^{j{\theta _{{\rm{SI}}{{\rm{M}}_l},k,1}}}}, \ldots ,{e^{j{\theta _{{\rm{SI}}{{\rm{M}}_l},k,N}}}}} \right) \in \mathbb{C}{^{N \times N}}$ denotes the phase-shift matrix where ${{e^{j{\theta _{{\rm{SI}}{{\rm{M}}_l},k,1}}}}}\in [-\pi, \pi ]$ represents the additional phase-shift of the LoS component between the $n$-th meta-atoms of SIM $l$ and UE $k$. This phase shift is induced by slight variations in the location of UE $k$ \cite{ozdogan2019performance}. Assuming uniformity across all elements of ${{\bf{\Theta }}_{{\rm{SI}}{{\rm{M}}_l},k}}$ the LoS component in \eqref{h_simk} can be expressed as ${\bf{\bar h}}_{{\rm{SI}}{{\rm{M}}_l},k}^{}{e^{j{\varphi _{{\rm{SI}}{{\rm{M}}_l},k}}}}$.\footnote{We assume a far-field scenario where the movement of UEs can be approximated as identical for each meta-atoms on the same SIM.} Then, the total channel ${{\bf{h}}_{l,k}} \in \mathbb{C}{^U}$ between AP $l$ and UE $k$ can be denoted as
\begin{align}\label{h_lk}
{{\bf{h}}_{l,k}} &= {\bf{W}}_{l,1}^{\rm{H}}{\bf{G}}_l^{\rm{H}}{\bf{h}}_{{\rm{SI}}{{\rm{M}}_l},k}^{}\notag\\
 &= {\bf{W}}_{l,1}^{\rm{H}}{\bf{G}}_l^{\rm{H}}{\bf{\bar h}}_{{\rm{SI}}{{\rm{M}}_l},k}^{}{e^{j{\varphi _{{\rm{SI}}{{\rm{M}}_l},k}}}} + {\bf{W}}_{l,1}^{\rm{H}}{\bf{G}}_l^{\rm{H}}{\bf{g}}_{{\rm{SI}}{{\rm{M}}_l},k}^{}.
\end{align}
\begin{rem}
We notice that in \eqref{h_lk}, the SIM alters the state of the channel by adjusting the phase shifts within $\mathbf{G}_l$. Unlike the direct channels in classical MIMO systems, the channel here is influenced by signals aggregated from multiple metasurface layers of SIM. 
\end{rem}

\subsection{Uplink Pilot Training and Channel Estimation}
In this section, it is posited that each UE sends pilot signals to all APs, which then use these incoming signals to estimate the aggregated channel. Specifically, we utilize $\tau_p$ mutually orthogonal pilot sequences for channel estimation. Let ${\mathbf{\phi}}_k \in \mathbb{C} ^{\tau_p}$ denotes the pilot sequence of UE $k$, satisfying ${\left\| {\phi _k^{}} \right\|^2} = {\tau _p}$. We utilize $\mathcal{P}_k$ to represent the index subset of UEs that employ the same pilot sequence as UE $k$ which includes UE $k$ itself. It is commonly assumed that in CF systems, the number of UEs exceeds the number of pilots, thus resulting in two or more UEs sharing the same pilot. Then, the signal ${\bf{y}}_l^p \in \mathbb{C}{^{U \times {\tau _p}}}$ received at AP $l$ can be written as
\begin{align}\label{y_lp}
{\bf{y}}_l^p &= \sum\limits_{k = 1}^K {\sqrt {{{\hat p}_k}} } {{\bf{h}}_{l,k}}{\mathbf{\phi}} _k^{\rm{T}} + {\bf{n}}_l^p\notag\\
 &= \sum\limits_{k = 1}^K {\sqrt {{{\hat p}_k}} } {\bf{W}}_{l,1}^{\rm{H}}{\bf{G}}_l^{\rm{H}}{\bf{h}}_{{\rm{SI}}{{\rm{M}}_l},k}^{}{\mathbf{\phi}} _k^{\rm{T}} + {\bf{n}}_l^p,
\end{align}
where ${{\hat p}_k}$ denotes the pilot transmit power of UE $k$. Additionally, ${\mathbf{n}}^{p}_l \in \mathbb{C}^{U \times {\tau_p}}$ denotes the additive noise, with entries independently following $\mathcal{CN}(0,\sigma^2)$, where $\sigma^2$ denotes the noise power. To estimate channel $\mathbf{h}_{l,k}$, the received signal is multiplied with pilot sequence of UE $k$ to obtain ${\bf{y}}_{l,k}^p = {\bf{y}}_l^p\phi _k^ *  \in \mathbb{C}{^U}$ as 
\begin{align}\label{y_lkp}
{\bf{y}}_{l,k}^p &= \sqrt {{{\hat p}_k}} {\tau _p}{\bf{W}}_{l,1}^{\rm{H}}{\bf{G}}_l^{\rm{H}}{\bf{h}}_{{\rm{SI}}{{\rm{M}}_l},k} \notag\\
&+ \sum\limits_{j \in {{\cal P}_k}\backslash k}^K {\sqrt {{{\hat p}_l}} } {\bf{W}}_{l,1}^{\rm{H}}{\bf{G}}_l^{\rm{H}}{\bf{h}}_{{\rm{SI}}{{\rm{M}}_l},j}^{}\mathbf{\phi} _j^{\rm{T}}\mathbf{\phi} _k^ *  + {\bf{n}}_l^p\mathbf{\phi} _k^ * .
\end{align}
Different channel estimation techniques can be employed to acquire the estimated aggregated channel. Here, we consider the phase-aware MMSE estimator, which assumes that ${\bf{\bar h}}_{{\rm{SI}}{{\rm{M}}_l},k}^{}$, ${{\varphi _{{\rm{SI}}{{\rm{M}}_l},k}}}$ and ${\beta _{{\rm{SI}}{{\rm{M}}_l},k}}$ are available for AP $l$ \cite{bjornson2019making}. The MMSE channel estimator is designed to reduce channel estimation errors, thereby achieving commendable system performance. Then, we can derive the effective estimated channel ${{\bf{\hat h}}_{l,k}}$ as 
\begin{align}\label{hath_lk}
{{\bf{\hat h}}_{l,k}} \!=\!\! {\bf{W}}_{l,1}^{\rm{H}}{\bf{G}}_l^{\rm{H}}{\bf{\bar h}}_{{\rm{SI}}{{\rm{M}}_l},k}^{}{e^{j{\varphi _{{\rm{SI}}{{\rm{M}}_l},k}}}} \!+\! \sqrt {{{\hat p}_k}} {{\bf{R}}_{l,k}}{\bf{\Psi }}_{l,k}^{ - 1}\!\left( {{\bf{y}}_{l,k}^p \!\!-\!\! {\bf{\bar y}}_{l,k}^p} \!\right),
\end{align}
where ${\bf{\bar y}}_{l,k}^p = \sum\nolimits_{j \in {{\cal P}_k}} {\sqrt {{{\hat p}_j}} } {\tau _p}{\bf{W}}_{l,1}^{\rm{H}}{\bf{G}}_l^{\rm{H}}{\bf{\bar h}}_{{\rm{SI}}{{\rm{M}}_l},j}{e^{j{\varphi _{{\rm{SI}}{{\rm{M}}_l},j}}}}$ and ${\bf{\Psi }}_{l,k}^{ - 1} = \sum\nolimits_{j \in {{\cal P}_k}} {{{\hat p}_j}} {\tau _p}{{\bf{R}}_{l,j}} + {\sigma ^2}{{\bf{I}}_{{U}}}$, where ${{\bf{R}}_{l,k}} = {\bf{W}}_{l,1}^{\rm{H}}{\bf{G}}_l^{\rm{H}}{{\bf{R}}_{{\rm{SI}}{{\rm{M}}_l},k}}{\bf{G}}_l{\bf{W}}_{l,1}$. The estimated channel ${{\bf{\hat h}}_{l,k}}$ and the estimation error ${{{\bf{\tilde h}}}_{l,k}} = {{\bf{h}}_{l,k}} - {{{\bf{\hat h}}}_{l,k}}$ are independent random variables with 
\begin{align}\label{var}
&\mathbb{E}\left\{ {{{{\bf{\hat h}}}_{l,k}}\left| {{{\varphi _{{\rm{SI}}{{\rm{M}}_l},k}}}} \right.} \right\} = {{{\bf{\bar h}}}_{l,k}}{e^{j{{{\varphi _{{\rm{SI}}{{\rm{M}}_l},k}}}}}},  \mathbb{E}\left\{ {{{{\bf{\tilde h}}}_{l,k}}} \right\} = {\bf{0}}, \notag\\
&{\rm{Cov}}\left\{ {{{{\bf{\hat h}}}_{l,k}}\left| {{{{\varphi _{{\rm{SI}}{{\rm{M}}_l},k}}}}} \right.} \right\} = {{\hat p}_k}{\tau _p}{{\bf{\Omega }}_{l,k}},\:{\rm{Cov}}\left\{ {{{{\bf{\tilde h}}}_{l,k}}} \right\} = {{\bf{C}}_{l,k}}, 
\end{align}
where ${\bf{\bar h}}_{l,k}^{} = {\bf{W}}_{l,1}^{\rm{H}}{\bf{G}}_l^{\rm{H}}{\bf{\bar h}}_{{\rm{SI}}{{\rm{M}}_l},k}^{}$, ${{\bf{\Omega }}_{l,k}} = {{\bf{R}}_{l,k}}{\bf{\Psi }}_{l,k}^{ - 1}{{\bf{R}}_{l,k}}$, and ${{\bf{C}}_{l,k}} = {{\bf{R}}_{l,k}} - {\hat p_k}{\tau _p}{{\bf{R}}_{l,k}}{\bf{\Psi }}_{l,k}^{ - 1}{{\bf{R}}_{l,k}}$. 
\begin{rem}
Note that during the channel estimation phase, it is customary to stabilize the SIM hardware and phase shifts to guarantee estimation precision. Given this premise, it is posited that both $\mathbf{W}_{l,1}$ and $\mathbf{G}_l$ in \eqref{hath_lk} remain constant and are known a priori.
\end{rem}

\newcounter{mytempeqncnt}
\begin{figure*}[t!]
\normalsize
\setcounter{mytempeqncnt}{1}
\setcounter{equation}{12}
\begin{align}\label{SINR_k}
{\gamma _k} = \frac{{{p_k}{{\left| \mathbb{E}{\left\{ {\sum\limits_{l = 1}^L {a_{l,k}^ * } {\bf{v}}_{l,k}^{\rm{H}}{\bf{W}}_{l,1}^{\rm{H}}{\bf{G}}_l^{\rm{H}}{\bf{h}}_{{\rm{SI}}{{\rm{M}}_l},k}^{}} \right\}} \right|}^2}}}{{\sum\limits_{j = 1}^K {{p_j}\mathbb{E}\left\{ {{{\left| {\sum\limits_{l = 1}^L {a_{l,k}^ * {\bf{v}}_{l,k}^{\rm{H}}{\bf{W}}_{l,1}^{\rm{H}}{\bf{G}}_l^{\rm{H}}{\bf{h}}_{{\rm{SI}}{{\rm{M}}_l},j}^{}} } \right|}^2}} \right\}}  - {p_k}{{\left| \mathbb{E}{\left\{ {\sum\limits_{l = 1}^L {a_{l,k}^ * } {\bf{v}}_{l,k}^{\rm{H}}{\bf{W}}_{l,1}^{\rm{H}}{\bf{G}}_l^{\rm{H}}{\bf{h}}_{{\rm{SI}}{{\rm{M}}_l},k}^{}} \right\}} \right|}^2} + {\sigma ^2}\mathbb{E}\left\{ {{{\left\| {\sum\limits_{l = 1}^L {a_{l,k}^ * } {\bf{v}}_{l,k}^{\rm{H}}} \right\|}^2}} \right\}}}.
\end{align}
\setcounter{equation}{13}
\hrulefill
\end{figure*}

\begin{figure*}[t!]
\normalsize
\setcounter{mytempeqncnt}{1}
\setcounter{equation}{13}
\begin{align}\label{SINR_k_close}
{\gamma _k} = \frac{{{p_k}{\bf{a}}_k^{\rm{H}}{{\bf{z}}_k}{\bf{z}}_k^{\rm{H}}{{\bf{a}}_k}}}{{\sum\limits_{j = 1}^K {{p_j}} {\bf{a}}_k^{\rm{H}}{{\bf{\Xi }}_{k,j}}{{\bf{a}}_k} + \sum\limits_{j \in {{\cal P}_k}\backslash \left\{ k \right\}} {{p_j}{{\hat p}_k}} {{\hat p}_j}\tau _p^2{\bf{a}}_k^{\rm{H}}{{\bf{\Delta }}_{k,j}}{\bf{\Delta }}_{k,j}^{\rm{H}}{{\bf{a}}_k} - {p_k}{\bf{a}}_k^{\rm{H}}{\bf{\Lambda }}_k^{}{\bf{\Lambda }}_k^{\rm{H}}{{\bf{a}}_k} + {\sigma ^2}{\bf{a}}_k^{\rm{H}}{{\bf{\Gamma }}_k}{\bf{a}}_k^{\rm{H}}}}.
\end{align}
\setcounter{equation}{14}
\hrulefill
\end{figure*}

\subsection{Uplink Data Transmission}
In the uplink, we assume that all UEs simultaneously transmit the uplink data symbols to all APs within each coherence block. The received signal ${\bf{y}}_l \in \mathbb{C}{^U}$ at AP $l$ is given by 
\begin{align}\label{y_l}
\setcounter{equation}{8}
{\bf{y}}_l = \sum\limits_{k = 1}^K {{{\bf{h}}_{l,k}}{s_k}}  + {\bf{n}}_l = \sum\limits_{k = 1}^K {{\bf{W}}_{l,1}^{\rm{H}}{\bf{G}}_l^{\rm{H}}{\bf{h}}_{{\rm{SI}}{{\rm{M}}_l},k}^{}{s_k}}  + {\bf{n}}_l,
\end{align}
where $s_k^{} \sim {\cal C}{\cal N}\left( {0,{p_k}} \right)$ represents the uplink transmitted signal by UE $k$ with power ${p_k} = \mathbb{E} \{ {{{\left| {{s_k}} \right|}^2}} \}$, and ${\bf{n}}_l^{} \sim {\cal C}{\cal N}\left( {0,{\sigma ^2}{{\bf{I}}_U}} \right)$ represents the additive noise. 

To fully leverage the characteristics of the multi-layer architecture of CF mMIMO systems, we assume that each AP performs preliminary signal processing locally, then transmits the processed signal to the CPU, where centralized signal merging and decoding are carried out. Let ${{\bf{v}}_{l,k}}$ denote the combining vector. Without losing generality, ${{\bf{v}}_{l,k}}$ is designed by AP $l$ and the local detect signal ${{\mathord{\buildrel{\lower3pt\hbox{$\scriptscriptstyle\smile$}} \over s} }_{l,k}}$ is derive as
\begin{align}\label{s_AP}
&{{\mathord{\buildrel{\lower3pt\hbox{$\scriptscriptstyle\smile$}} 
\over s} }_{l,k}} = {\bf{v}}_{l,k}^{\rm{H}}\sum\limits_{k = 1}^K {{\bf{W}}_{l,1}^{\rm{H}}{\bf{G}}_l^{\rm{H}}{\bf{h}}_{{\rm{SI}}{{\rm{M}}_l},k}^{}{s_k}}  + {\bf{v}}_{l,k}^{\rm{H}}{\bf{n}}_l \\
&=\! {\bf{v}}_{l,k}^{\rm{H}}\!{\bf{W}}_{l,1}^{\rm{H}}\!{\bf{G}}_l^{\rm{H}}{\bf{h}}_{{\rm{SI}}{{\rm{M}}_l},k}^{}{s_k} \!+\! {\bf{v}}_{l,k}^{\rm{H}}\!\!\!\!\!\!\!\sum\limits_{j = 1,j \ne k}^K \!\!\!\!\!\! {{\bf{W}}_{l,1}^{\rm{H}}\!{\bf{G}}_l^{\rm{H}}{\bf{h}}_{{\rm{SI}}{{\rm{M}}_l},j}{s_j}} \! +\!\! {\bf{v}}_{l,k}^{\rm{H}}{\bf{n}}_l.\notag
\end{align}
Utilizing known local CSI, APs can adopt a variety of schemes to design the combining vector $\mathbf{v}_{l,k}$. We consider the MR combining methods \cite{ozdogan2019performance} where ${{\bf{v}}_{l,k}} = {{\bf{\hat h}}_{l,k}}$. 
To further reduce inter-user interference, the preliminary decoded signal $\left\{ {{{\mathord{\buildrel{\lower3pt\hbox{$\scriptscriptstyle\smile$}} 
\over s} }_{l,k}}:l = 1, \ldots ,L} \right\}$ at the AP are transmitted to the CPU via the fronthaul link, where they are multiplied by the weight coefficient ${a_{l,k} }$ for final decoding. Then, the derived signal ${{\hat s}_k} = \sum\limits_{l = 1}^L {a_{l,k}^ * } {{\mathord{\buildrel{\lower3pt\hbox{$\scriptscriptstyle\smile$}} 
\over s} }_{l,k}}$ can be obtained as
\begin{align}\label{s_CPU}
{{\hat s}_k} &= \sum\limits_{l = 1}^L {a_{l,k}^ * } {\bf{v}}_{l,k}^{\rm{H}}{\bf{W}}_{l,1}^{\rm{H}}{\bf{G}}_l^{\rm{H}}{\bf{h}}_{{\rm{SI}}{{\rm{M}}_l},k}^{}{s_k} \\
&+ \sum\limits_{j = 1,j \ne k}^K {\sum\limits_{l = 1}^L {a_{l,k}^ * {\bf{v}}_{l,k}^{\rm{H}}} {\bf{W}}_{l,1}^{\rm{H}}{\bf{G}}_l^{\rm{H}}{\bf{h}}_{{\rm{SI}}{{\rm{M}}_l},j}^{}{s_j}}  + \sum\limits_{l = 1}^L {a_{l,k}^ * } {\bf{v}}_{l,k}^{\rm{H}}{\bf{n}}_l.\notag
\end{align}

\section{Closed-form Expressions and Performance Analysis}
In this section, we explore the uplink SE performance of the SIM-enhanced CF mMIMO system with EGCD and LSFD. Then, we investigate the impact of SIM on system performance.

Based on \eqref{s_CPU}, the achievable SE lower bound of UE $k$ is obtained by adopting the use-and-then-forget (UatF) bound \cite{ozdogan2019performance} as 
\begin{align}\label{SE_k}
{\rm{S}}{{\rm{E}}_k} = \frac{{{\tau _u}}}{{{\tau _c}}}{\log _2}\left( {1 + {\gamma _k}} \right),
\end{align}
where the effective signal-to-interference-plus-noise ratio (SINR) $\gamma_k$ is provided as \eqref{SINR_k} at the top of this page. 

Various previous works have demonstrated that employing a two-layer signal processing architecture in CF mMIMO systems can achieve superior performance \cite{bjornson2020rayleigh,bjornson2017massive,zheng2021impact}. This involves conducting preliminary local signal processing at the APs, followed by the transmission of processed information to a CPU for centralized signal decoding. 
In this context, we consider adopting MR combining at the APs. At the CPU, we compare LSFD and EGCD to further enhance system performance and derive closed-form expressions for SE.

\subsubsection{Large Scale Fading Decoding (LSFD)} 
At the CPU, $\mathbf{a}_k$ can be optimized only by using the channel statistics information to maximize the SE. By adopting this approach, the CPU does not need to know all the instantaneous channel information, which reduces the fronthaul cost and the computational load. In \cite{van2018large}, this approach is known as LSFD in cellular massive MIMO which can be applied here.
Through mathematical derivation, we can arrive at the following Theorem~\ref{Them}.

\begin{them}\label{Them}
The closed-form expression for the uplink SE of UE $k$ can be expressed by \eqref{SE_k}, where the SINR with MR combining and LSFD is given by \eqref{SINR_k_close} at the top of this page. Specifically, the desired signal is denoted by the following formula 
\begin{align}\label{DS}
\setcounter{equation}{14}
&{\bf{a}}_k^{} = {\left[ {{a_{1,k}}, \ldots ,{a_{L,k}}} \right]^{\rm{T}}} \in \mathbb{C}{^L},\\
&{\left[ {{{\bf{z}}_k}} \right]_l} = {\rm{tr}}\left( {{p_k}{\tau _p}{{\bf{\Omega }}_{l,k}} + {{\left\| {{\bf{\bar h}}_{l,k}^{}} \right\|}^2}} \right),
\end{align}
where ${\bf{\bar h}}_{l,k}^{} = {\bf{W}}_{l,1}^{\rm{H}}{\bf{G}}_l^{\rm{H}}{\bf{\bar h}}_{{\rm{SI}}{{\rm{M}}_l},k}$. Moreover, the definition of non-coherent interference is given by
\begin{align}\label{NC_UI}
{{\bf{\Xi }}_{k,j}} &= {\rm{diag}}\left( {{\xi _{1,kj}}, \ldots ,{\xi _{L,kj}}} \right) \in \mathbb{C}{^{L \times L}},\\
{\xi _{l,kj}} &= {{\hat p}_k}{\tau _p}{\rm{tr}}\left( {{{\bf{R}}_{l,j}}{{\bf{\Omega }}_{l,k}}} \right) + {\bf{\bar h}}_{l,k}^{\rm{H}}{{\bf{R}}_{l,j}}{\bf{\bar h}}_{l,k}^{} \notag\\
&+ {{\hat p}_k}{\tau _p}{\bf{\bar h}}_{l,j}^{\rm{H}}{{\bf{\Omega }}_{l,k}}{\bf{\bar h}}_{l,j}^{} + {\left| {{\bf{\bar h}}_{l,k}^{\rm{H}}{\bf{\bar h}}_{l,j}^{}} \right|^2}.
\end{align}
The coherent interference is given by
\begin{align}\label{C_UI}
{{\bf{\Delta }}_{k,j}} \!\!&=\!\! {\left[ {{\rm{tr}}\!\left(\! {{{\bf{R}}_{1,j}}{\bf{\Psi }}_{1,k}^{ - 1}{{\bf{R}}_{1,k}}} \!\right)\!,\! \ldots \!,\!{\rm{tr}}\!\left(\! {{{\bf{R}}_{L,j}}{\bf{\Psi }}_{L,k}^{ - 1}{{\bf{R}}_{L,k}}} \!\right)}\! \right]^{\rm{T}}} \!\!\!\!\!\in \!\mathbb{C}{^L},\\
{\bf{\Lambda }}_k^{} &= {\left[ {{\rm{tr}}\left( {{{\left\| {{\bf{\bar h}}_{1,k}^{}} \right\|}^2}} \right), \ldots ,{\rm{tr}}\left( {{{\left\| {{\bf{\bar h}}_{L,k}^{}} \right\|}^2}} \right)} \right]^{\rm{T}}} \in \mathbb{C}{^L}.
\end{align}
Finally, the noise is denoted by ${{\bf{\Gamma }}_k} = {\rm{diag}}\left( {{{\bf{z}}_k}} \right) \in \mathbb{C}{^{L \times L}}$.
\end{them}

\begin{IEEEproof}
The proof of \eqref{SINR_k_close} is given in Appendix A.
\end{IEEEproof}

Furthermore, we utilize the LSFD to obtain $\mathbf{a}_k$ for maximizing the SINR in \eqref{SINR_k_close}, yielding \cite{shi2022spatially}
\begin{align}\label{a_LSFD}
    {\bf{a}}_k^{} &= \left( {\sum\limits_{j = 1}^K {{p_j}} {{\bf{\Xi }}_{k,j}}} \right. + \sum\limits_{j \in {{\cal P}_k}\backslash \left\{ k \right\}} {{p_j}{{\hat p}_k}} {{\hat p}_j}\tau _p^2{{\bf{\Delta }}_{k,j}}{\bf{\Delta }}_{k,j}^{\rm{H}} \notag\\
&{\left. { - {p_k}{\bf{\Lambda }}_k^{}{\bf{\Lambda }}_k^{\rm{H}} + {\sigma ^2}{{\bf{\Gamma }}_k}} \right)^{ - 1}}{{\bf{z}}_k}.
\end{align}
Consequently, the SINR in \eqref{SINR_k_close} can be derived as
\begin{align}\label{SINR_LSFD}
{\gamma _k} &= {p_k}{\bf{z}}_k^{\rm{H}}\left( {\sum\limits_{j = 1}^K {{p_j}} {{\bf{\Xi }}_{k,j}}} \right. + \sum\limits_{j \in {{\cal P}_k}\backslash \left\{ k \right\}} {{p_j}{{\hat p}_k}} {{\hat p}_j}\tau _p^2{{\bf{\Delta }}_{k,j}}{\bf{\Delta }}_{k,j}^{\rm{H}}\notag\\
&{\left. { - {p_k}{\bf{\Lambda }}_k^{}{\bf{\Lambda }}_k^{\rm{H}} + {\sigma ^2}{{\bf{\Gamma }}_k}} \right)^{ - 1}}{{\bf{z}}_k}.
\end{align}
\begin{IEEEproof}
The proof of \eqref{a_LSFD} and \eqref{SINR_LSFD} can be obtained following the same matrix derivation in \cite{bjornson2019making} and is therefore omitted.
\end{IEEEproof}

\subsubsection{Equal Gain Combining Decoding (EGCD)} 
In this case, the signals processed by the APs are transmitted to the CPU, where signals from different APs are combined with equal gain combining (EGC) for final decoding \cite{ikki2009performance}. This is a specific case of LSFD, with the decoding coefficient given by ${\bf{a}}_k^{} = {\left[ {1, \ldots ,1} \right]^{\rm{T}}} \in \mathbb{C}{^L}$.

\begin{rem}
Note that the aggregated channels and the spatial correlation via the SIM affect the system performance by influencing ${{\bf{R}}_{l,k}} = {\bf{W}}_{l,1}^{\rm{H}}{\bf{G}}_l^{\rm{H}}{{\bf{R}}_{{\rm{SI}}{{\rm{M}}_l},k}}{\bf{G}}_l{\bf{W}}_{l,1}$ in the closed-form \eqref{SINR_k_close}. By increasing the number of meta-atoms in each metasurface layer of the SIM, ${{\bf{R}}_{l,k}}$ is enhanced, which leads to a higher numerator in \eqref{SINR_k_close}. This improvement in the numerator contributes to better system performance. However, increasing the number of SIM metasurface layers results in a larger number of $W_{l,m}$ multiplications within ${\bf{G}}_l$, leading to more severe large-scale fading. Therefore, the phase shift configuration is more crucial in SIM-enhanced CF mMIMO systems for compensating the adverse path loss.
\end{rem}

\section{System Optimization and Design}
In this section, we utilize the derived closed-form SE expressions to design pilot allocation, SIM wave-based beamforming, and power control in the SIM-enhanced CF system, aiming to unlock the potential of the considered system and enhance system performance.

\subsection{Pilot Allocation}
In CF systems, the number of pilot resources is often less than the number of users, leading to pilot contamination among users \cite{liu2020graph}. Therefore, it is necessary to design a reasonable pilot allocation scheme to reduce inter-user interference. Here, we propose a pilot allocation algorithm to minimize inter-user interference as shown in Algorithm 1. 
First, pilots are randomly assigned to the first $\tau_p$ UEs as part of the initialization. Subsequently, the algorithm iteratively allocates pilots to the remaining UEs. In each iteration, the inter-user interference (UI) for each potential pilot choice is calculated based on the current pilot set. The pilot resulting in the least interference is selected and allocated to the respective UE. This process is repeated until pilots have been assigned to all UEs.

\begin{algorithm}[t]
\caption{Pilot Allocation Algorithm}
\begin{algorithmic}[1]
\State \textbf{Input:} $\mathbf{R}_{l,k}$, ${\bf{\bar h}}_{{\rm{SI}}{{\rm{M}}_l},k}$, $L$, $K$, $U$, $\tau_p$.
\State \textbf{Initialize:} Pilots are randomly assigned to the first $\tau_p$ UEs, $\mathcal{P}_{set} \gets [1, 2, \dots, \tau_p]$.
\For{$z = 1$ \textbf{to} $(K/\tau_p) - 1$}
    \State $\mathcal{P}_{set} \gets [\mathcal{P}_{set};\, ((\tau_p \times z) + 1) \times \rm{ones}(1, \tau_p)]$, $ind \gets [\;]$;
    \For{$t = 1$ \textbf{to} $\tau_p$}
        \State Calculate the inter-user interference (UI) for each 
        \Statex \hspace{\algorithmicindent}\;\;\;\; potential pilot choice based on the current pilot
        \Statex \hspace{\algorithmicindent}\;\;\;\; set;
        \State $[ \sim, ind(t)] \gets \min(\rm{UI})$, select the pilot with the
        \Statex \hspace{\algorithmicindent}\;\;\;\; minimum interference for the current UE;
        \State \, $x \gets [1, 2, \dots, \tau_p]$
        \State \textbf{remove} $ind$ \textbf{from} $x$
        \State ${\mathcal{P}_{set}}(z+1, x) \gets (z \times \tau_p) + t + 1$;
    \EndFor
\EndFor
\State Order the pilot allocation set;
\State \textbf{Output:} $\mathcal{P}_{k} \gets \mathcal{P}_{set}$.
\end{algorithmic}
\end{algorithm}

\begin{algorithm}[t]
\caption{The SIM Wave-based Beamforming Algorithm for Solving \eqref{Phaseshift}}
\begin{algorithmic}[1]
\State \textbf{Input:} ${{\bf{W}}_{l,m}}$, ${{{\bf{R}}_{{\rm{SI}}{{\rm{M}}_l},k}}}$, ${\bf{\bar h}}_{{\rm{SI}}{{\rm{M}}_l},k}$, $L$, $U$, $N$, $M$, $p_k$.
\State \textbf{Initialize:} SIM phase shifts with random values in $[0, 2\pi)$ for dimensions $N \times M \times L$;
\State Compute initial SE;
\State Generate a random permutation of indices for all meta-atoms (${N_{all}} = N \times M \times L$);
\For{$l = 1$ \textbf{to} $L$}
    \For{$i = 1$ \textbf{to} ${N_{all}}$ \textbf{step} $N_{selection}$}
    \State Select a subset of $N_{selection}$ indices starting from 
    \Statex \hspace{\algorithmicindent}\;\;\;\; the index $i$;
    \State Convert linear indices to subscript indices; (\textit{subN}, 
    \Statex \hspace{\algorithmicindent}\;\;\;\; \textit{subM}, \textit{subL}) corresponding to dimensions of SIM 
    \Statex \hspace{\algorithmicindent}\;\;\;\; phase shifts;
    \State Compute \textit{current\_SE} by using \eqref{SINR_k_close} and initialize 
    \Statex \hspace{\algorithmicindent}\;\;\;\; \textit{best\_SE} to \textit{current\_SE};
    \State Initialize phase shift to the phases at indices 
    \Statex \hspace{\algorithmicindent}\;\;\;\; (\textit{subN, subM, subL});
        \For{$j = 1$ \textbf{to} $J$}
            \State Initialize $N_{selection}$ new phase shifts as a copy 
            \Statex \hspace{\algorithmicindent}\hspace{\algorithmicindent}\;\;\;\; of SIM phase shifts, the $N_{selection}$ new phase 
            \Statex \hspace{\algorithmicindent}\hspace{\algorithmicindent}\;\;\;\; shifts are equal to the current phase shifts plus 
            \Statex \hspace{\algorithmicindent}\hspace{\algorithmicindent}\;\;\;\; a pre-set step size value;
            \State Compute \textit{new\_SE} using the updated SIM 
            \Statex \hspace{\algorithmicindent}\hspace{\algorithmicindent}\;\;\;\; phase shifts;
            \If{\textit{new\_SE} $>$ \textit{best\_SE} $+$ $\xi$ }
                \State Update \textit{best\_SE} to \textit{new\_SE};
                \State Update phase shifts to new phase shifts at
                \Statex \hspace{\algorithmicindent}\hspace{\algorithmicindent}\hspace{\algorithmicindent}\;\;\;\; (\textit{subN}, \textit{subM}, \textit{subL});
                \State \textbf{break}
            \EndIf
        \EndFor
    \EndFor
\State Update the all phase shifts of SIM at (\textit{subN}, \textit{subM}, 
\Statex \hspace{\algorithmicindent} \textit{subL});
\State Update \textit{current\_SE} to \textit{best\_SE};
\State \textbf{Output:} Final SE and SIM wave-based beamforming coefficient;
\EndFor
\end{algorithmic}
\end{algorithm}

\subsection{SIM Wave-based Beamforming Design}
In this section, we propose a SIM wave-based beamforming optimization algorithm aimed at maximizing the sum SE of the SIM-enhanced CF mMIMO system. The optimization problem can be modeled as follows:
\setcounter{equation}{22}
\begin{subequations}\label{Phaseshift}
  \begin{align}
  &\mathop {{\rm{max}}}\limits_{\varphi _{l,m}^n} \;\;\sum\limits_{k = 1}^K {{\gamma _k}} \label{eq:first}\\
  &\;{\rm{s.}}\;{\rm{t.}}\quad \varphi _{l,m}^n \in \left[ {0,2\pi } \right),\;\forall l \in {\cal L},\;\forall m \in {\cal M},\;\forall n \in {\cal N}. \label{eq:second}
  \end{align}
\end{subequations}
The problem \eqref{Phaseshift} is a complex non-convex issue, and optimizing a large number of SIM meta-atoms simultaneously would result in extremely high complexity. By fully considering the multi-layer structure characteristic of SIM, we adopt the concept of iterative optimization, sequentially optimizing the meta-atoms of each layer of SIM as shown in Algorithm 2. 
Specifically, the algorithm is divided into two alternating iterative processes: cyclic iteration for SIM and cyclic iteration for layers within a SIM. Initially, a target SIM is selected, followed by an optimization process for each layer within the current SIM. Each iteration involves adding a predefined phase step to the meta-atoms of that layer, followed by an evaluation to determine if the SE has improved. If there is an improvement, the current phase shift is updated, and the iteration continues until the target gain $\xi$ is achieved or the maximum number of cycles $J$ is reached. The process then repeats for the next layer until all layers of the current SIM have been optimized. Subsequently, the next SIM is selected, and the aforementioned steps are executed.
The computational complexity associated with Algorithm 2 is quantified as $\mathcal{O}(NML) + \frac{NML}{N_{\text{selection}}} \cdot J \cdot \mathcal{O}(f(\rm{SE}))$, where \( f(\rm{SE}) \) is the function representing the time complexity to compute the SE. It is clear that increasing the number of SIM meta-atoms will significantly increase computational complexity. Also, by reducing the number of inner loop iterations $J$ or increasing the number of meta-atoms $N_{\rm{selection}}$ selected each time, computational complexity can be lowered at the expense of performance.
\begin{rem}
    Note that this iterative algorithm only guarantees a locally optimal solution rather than the globally optimal solution \cite{you2020energy}. In practice, the approach of narrowing the step size and increasing the number of iterations can be used to approximate the best possible beamforming outcome at the cost of increased complexity. In this paper, we focus on the analysis of the effects brought by the integration of SIM with the CF mMIMO system. Therefore, solving for the optimal wave-based beamforming is not the primary focus of this study.
\end{rem}
\begin{rem}
    %The algorithm we propose requires only statistical CSI rather than instantaneous CSI, which significantly reduces the load on the fronthaul link. This also means that SIM needs to perform wave-based beamforming design only on a large time scale, thereby decreasing the frequency of computation.
    The algorithm we propose operates using only statistical CSI instead of relying on instantaneous CSI. This significantly reduces the data load on the fronthaul link, as it eliminates the need for continuous and frequent updates of CSI. Additionally, this approach allows the SIMs to conduct wave-based beamforming design on a larger time scale, which in turn decreases the frequency of computational tasks. Thus, the overall system efficiency is enhanced by reducing the computational burden in processing instantaneous CSI.
\end{rem}

\begin{algorithm}[t]
\caption{Power Control Algorithm for Solving \eqref{maxmin2} } 
\begin{algorithmic}[1]
\State \textbf{Input:} ${\bf{a}}_k$, ${\bf{z}}_k$, ${{{\bf{\Xi }}_{k,j}}}$, ${{{\bf{\Delta }}_{k,j}}}$, ${{\bf{\Lambda }}_k}$, ${{{\bf{\Gamma }}_k}}$, $\tau_p$.
\State \textbf{Initialize:} Select the initial values for ${t_{\min}}$ and ${t_{\max}}$, where ${t_{\min}}$ and ${t_{\max}}$ establish a range of relevant values of the objective function in \eqref{maxmin2}. Additionally, choose a tolerance $\varepsilon  > 0$.      
\State Set $t: = \frac{{{t_{\min }} + {t_{\max }}}}{2}$. Proceed by addressing the subsequent convex feasibility problem:
\begin{align}\label{suanfa}
\left\{ {\begin{array}{*{20}{c}}
  {\frac{1}{t}{p_k}({{\bf{a}}_k^{\rm{H}}{{\bf{z}}_k}{\bf{z}}_k^{\rm{H}}{{\bf{a}}_k}}) \geqslant {\zeta _k},\;k = 1, \ldots K,} \\
  {\;0 \leqslant {p_k} \leqslant {p_{\max }},\;k = 1, \ldots K,}
\end{array}} \right.
\end{align}
where ${\zeta _k}\! =\!\! {\sum\limits_{j = 1}^K {{p_j}} {\bf{a}}_k^{\rm{H}}{{\bf{\Xi }}_{k,j}}{{\bf{a}}_k} \!+ \!\!\!\!\!\!\!\!\!\!\sum\limits_{j \in {{\cal P}_k}\backslash \left\{ k \right\}} \!\!\!\!\!\!\!{{p_j}{{\hat p}_k}} {{\hat p}_j}\tau _p^2{\bf{a}}_k^{\rm{H}}{{\bf{\Delta }}_{k,j}}{\bf{\Delta }}_{k,j}^{\rm{H}}{{\bf{a}}_k}}$
${ - {p_k}{\bf{a}}_k^{\rm{H}}{\bf{\Lambda }}_k^{}{\bf{\Lambda }}_k^{\rm{H}}{{\bf{a}}_k} + {\sigma ^2}{\bf{a}}_k^{\rm{H}}{{\bf{\Gamma }}_k}{\bf{a}}_k^{\rm{H}}}$;
\State If problem \eqref{suanfa} is feasible, then set ${t_{\min }}: = t$, else set ${t_{\max }}: = t$;
\State Stop if ${t_{\max }} - {t_{\min }} < \varepsilon $. Otherwise, go to step 2;
\State Compute final SE;
\State \textbf{Output:} Final power coefficient and SE.
\end{algorithmic}
\end{algorithm}

\subsection{Power Control Design}
The original intent of CF mMIMO systems is to achieve more uniform coverage of all UEs in the scenario through a large number of distributed APs deployment, thereby improving the performance of UEs with poor channel conditions \cite{ngo2017cell}. Therefore, in this section, we focus on a power control algorithm designed to improve the service quality for users with inferior performance. Based on that, we propose the max-min SE power control method based on the closed-form expression in \eqref{SINR_k_close}, and the optimization problem can be formulated as
\begin{subequations}\label{maxmin1}
  \begin{align}
  &\mathop {{\text{max}}}\limits_{{p_k}} \; \mathop {{\text{min}}}\limits_{k = 1, \cdots ,K} \; \gamma _k \label{eq:first1} \\
  &\;{\text{s.}}\;{\text{t.}}\quad\quad\quad\quad\;\; 0 \leqslant {p_k} \leqslant {p_{\max }},\; \forall k \in {\cal K}, \label{eq:second1}
  \end{align}
\end{subequations}
where $\gamma _k$ is given by \eqref{SINR_k_close}. By introducing an auxiliary optimization variable $t$, problem \eqref{maxmin1} can be equivalently reformulated as
\begin{subequations}\label{maxmin2}
  \begin{align}
  &\mathop {{\text{max}}}\limits_{{p_k},t} \quad t \label{eq:first2}  \\
  &\;{\text{s.}}\;{\text{t.}}\quad\; t \leqslant \gamma _k,\;\forall k \in {\cal K}, \label{eq:second2}\\
  &\quad \quad \;\;\; 0 \leqslant {p_k} \leqslant {p_{\max }},\;\forall k \in {\cal K}. \label{eq:third2}
  \end{align}
\end{subequations}
%Consequently, problem \eqref{maxmin2} can be efficiently solved by using bisection and solving a sequence of linear feasibility problems as shown in Algorithm 3. Note that in Algorithm 3, if we focus on a reasonable range of the initial upper and lower bounds (i.e., $t_{\min} = 0$ and ${t_{\max }} \geqslant \max \left[ {{\gamma _k}} \right]$), we can effectively find a feasible solution within a few iterations. The computational complexity of Algorithm 3 is $\mathcal{O}\left( {2K^2+2K} \right)$ and it is clear that complexity increases rapidly as the number of users increases. As such, in practice, when the number of UEs is small, exploiting the max-min method is recommended.
Consequently, problem \eqref{maxmin2} can be addressed effectively using a bisection method and tackling a series of linear feasibility challenges, as detailed in Algorithm 3. It's important to note that, within Algorithm 3, by setting appropriate initial bounds for the bisection method ($t_{\min} = 0$ and ${t_{\max }} \geqslant \max \left[ {{\gamma _k}} \right]$), a viable solution can usually be identified in just a few iterations. The computational complexity associated with Algorithm 3 is quantified as $\mathcal{O}\left( {2K^2+2K} \right)$, highlighting that the complexity escalates quickly with an increase in the number of users. Therefore, in scenarios where the number of UEs is small, it is advantageous to utilize the max-min method.

\section{Numerical Results and Discussion}
\subsection{Simulation Setup}
This section presents numerical results and discussion to validate the accuracy of the derived analysis and to evaluate the performance of the SIM-enhanced CF mMIMO system.
We assume that the APs and UEs are uniformly distributed in the $500 \times 500 \:\rm{m}^2$ area with a wrap-around scheme \cite{bjornson2019making}. Furthermore, a SIM, which stacks multiple metasurfaces, is integrated into the AP to execute transmit beamforming in the EM wave domain. The height of the SIM-enhanced APs and UEs is 15 $\rm{m}$ and 1.65 $\rm{m}$, respectively. Furthermore, we assume the thickness of the SIM is $T_{\rm{SIM}} = 5\lambda$, ensuring that the spacing between two adjacent metasurfaces in the $M$-layer SIM is ${d_{{\rm{Layer}}}} = {{{T_{{\rm{SIM}}}}} \mathord{\left/
 {\vphantom {{{T_{{\rm{SIM}}}}} M}} \right.
 \kern-\nulldelimiterspace} M}$. Furthermore, we consider a square metasurface structure SIM with $N={N_x}{N_y}$ meta-atoms and $N_x = N_y$ where $N_x$ and $N_y$ denotes the number of meta-atoms along the $x$-axis and $y$-axis, respectively. Furthermore, we assume half-wavelength spacing between adjacent antennas/meta-atoms at the APs and metasurfaces. Then, the size of SIM meta-atom is $d = {d_x} = {d_y} = {\lambda  \mathord{\left/
 {\vphantom {\lambda  2}} \right.
 \kern-\nulldelimiterspace} 2}$. Moreover, we assume a half-wavelength spacing between meta-atoms on SIMs and antennas at the APs. 
 The COST-321 Walfish-Ikegami model \cite{bjornson2020rayleigh} is adopted to compute the pathloss as
\begin{align}
{\beta _{lk}}\left[ {{\rm{dB}}} \right] =  - 30.18 - 26{\log _{10}}\left( {\frac{{{d_{lk}}}}{{1\:{\rm{m}}}}} \right) + {F_{lk}},
\end{align}
where $d_{lk}$ denotes the distance between AP $l$ and UE $k$. The Rician $\kappa$-factor is denoted as ${\kappa _{lk}} = {10^{1.3 - 0.003{d_{lk}}}}$.
The shadow fading $F_{lk}$ and other parameters is similar to \cite{ngo2017cell} with ${F_{lk}} = \sqrt {{\delta _f}} {a_l} + \sqrt {1 - {\delta _f}} {b_{k}}$, where ${a_l} \sim {\cal N}\left( {0,\delta _{{\text{sf}}}^2} \right)$ and ${b_{k}} \sim {\cal N}\left( {0,\delta _{{\text{sf}}}^2} \right)$ are independent random variables and ${{\delta _f}}$ is the shadow fading parameter. The variances of $a_l$ and $b_{k}$ are $\mathbb{E}\left\{ {{a_l}{a_{l'}}} \right\} = {2^{ - \frac{{{d_{ll'}}}}{{{d_{{\text{dc}}}}}}}}$, $\mathbb{E}\left\{ {{b_{k}}{b_{k'}}} \right\} = {2^{ - \frac{{{d_{kk'}}}}{{{d_{{\text{dc}}}}}}}}$, respectively, where ${{d_{ll'}}}$ and ${{d_{kk'}}}$ are the geographical distances between AP $l$-AP $l'$ and UE $k$-UE $k'$, respectively. Moreover, $d_{\rm{dc}}$ is the decorrelation distance depending on the environment. We set ${{\delta _f}} = 0.5$, $d_{\rm{dc}} = 100$ m and ${\delta _{{\text{sf}}}} = 8$ in this paper. The large-scale coefficients of ${{\bf{H}}_m}$ are given by
\begin{align}
\beta _{lk}^{{\rm{LoS}}} = \frac{{{\kappa _{lk}}}}{{{\kappa _{lk}} + 1}}{\beta _{lk}},{\kern 1pt} \;\;\beta _{lk}^{{\rm{NLoS}}} = \frac{1}{{{\kappa _{lk}} + 1}}{\beta _{lk}}.
\end{align}
Each UE transmits with maxmize power $200\; \rm{mW}$ and we assume that the noise power is $\sigma^2 = -94\; \rm{dBm}$. For the block, we assume the length of the coherence block is $\tau_c = 200$ and $\tau_p = 4$ is used for pilot transmission.

\subsection{Impact of Parameters in CF mMIMO Systems}

\begin{figure}[t]
\centering
\includegraphics[scale=0.55]{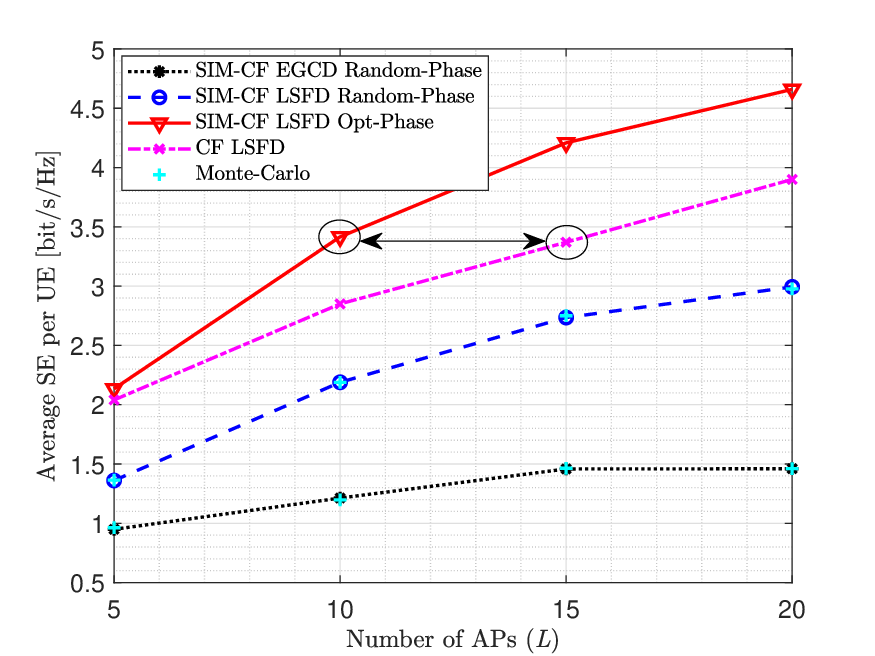}\vspace{-0.2cm}
\caption{Average SE per UE against different numbers of APs under EGCD/LSFD with AP full power transmission ($M = 5$, $N = 64$, $K = 5$, $U = 2$, $\tau_p = 4$, ${d_x} = {d_y} = {1 \mathord{\left/
 {\vphantom {1 2}} \right.
 \kern-\nulldelimiterspace} 2}\lambda$).}\label{Fig_L}\vspace{-0.4cm}
\end{figure}
Fig.~\ref{Fig_L} illustrates the average SE per UE against different numbers of APs under EGCD/LSFD with AP full power transmission. The results indicate that increasing the number of APs leads to an improvement in system performance. Also, compared with EGCD, adopting LSFD can achieve a twofold increase in system performance. It is clear that by adopting random phase shifts, the performance of the SIM-enhanced CF mMIMO system is inferior to that of the CF mMIMO system. This is due to the additional propagation losses introduced by the multi-layer metasurfaces of the SIM, and the random phase shifts fail to harness the potential of the numerous meta-atoms in SIM. However, by employing the proposed wave-based beamforming design, a significant enhancement in system performance can be achieved, surpassing that of the CF mMIMO system, and the gain gap further increases with the number of APs increase. In particular, with the introduction of SIM, deploying only 10 APs in the SIM-enhanced CF mMIMO system achieves SE performance that surpasses that of a traditional CF mMIMO system with 15 APs. This indicates that the introduction of SIMs can reduce the number of APs required in CF mMIMO systems, thereby saving costs.
%This reveals that SIM needs to be combined with advanced wave-based beamforming design schemes to unleash its potential for communication enhancement.

\begin{figure}[t]
\centering
\includegraphics[scale=0.55]{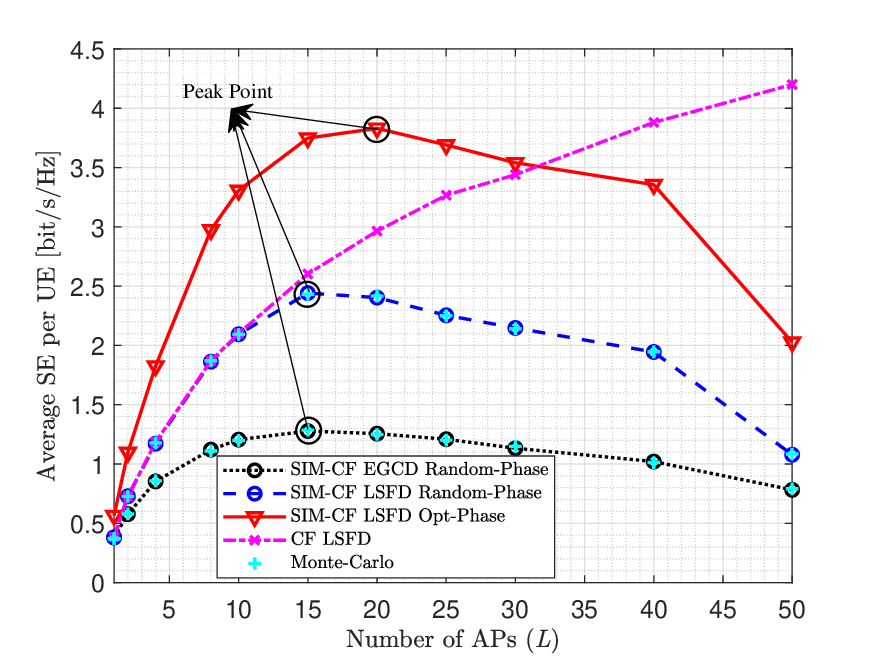}\vspace{-0.2cm}
\caption{Average SE per UE against different numbers of APs with the total number of meta-atoms $N_{\rm{total}} = 1200$ on all SIM-enhanced APs over EGCD/LSFD with AP full power transmission ($M = 5$, $K = 5$, $U = 1$, $\tau_p = 4$, ${d_x} = {d_y} = {1 \mathord{\left/
 {\vphantom {1 2}} \right.
 \kern-\nulldelimiterspace} 2}\lambda$).}\label{Fig_meta_unchange}\vspace{-0.4cm}
\end{figure}
Fig.~\ref{Fig_meta_unchange} illustrates the average SE per UE against different numbers of APs with the given total number of meta-atoms on all SIM over EGCD/LSFD with AP full power transmission. We assume that the total number of meta-atoms on all SIMs is $N_{\rm{total}} = L \times M \times N = 1200$. This is to investigate whether it is better to deploy a large number of small-sized SIM-enhanced APs or a smaller number of large-sized SIM-enhanced APs in the considered system. It is clear that as the number of APs increases, the performance of the SIM-enhanced CF mMIMO system first improves and then declines, exhibiting a peak. Moreover, when the number of APs $L$ exceeds 30, even with wave-based beamforming optimization of the SIM, the system performance still cannot surpass that of the CF mMIMO system. This reveals that in practical deployments, the number of meta-atoms per SIM should not be too small; otherwise, system performance may be difficult to improve, even with a large deployment of SIMs. Additionally, the peak performance points for optimized and random phase shifts occur at $L = 20$ and $L = 15$, respectively, indicating that wave-based beamforming optimization can increase the optimal number of APs that the system can accommodate.

\begin{figure}[t]
\centering
\includegraphics[scale=0.55]{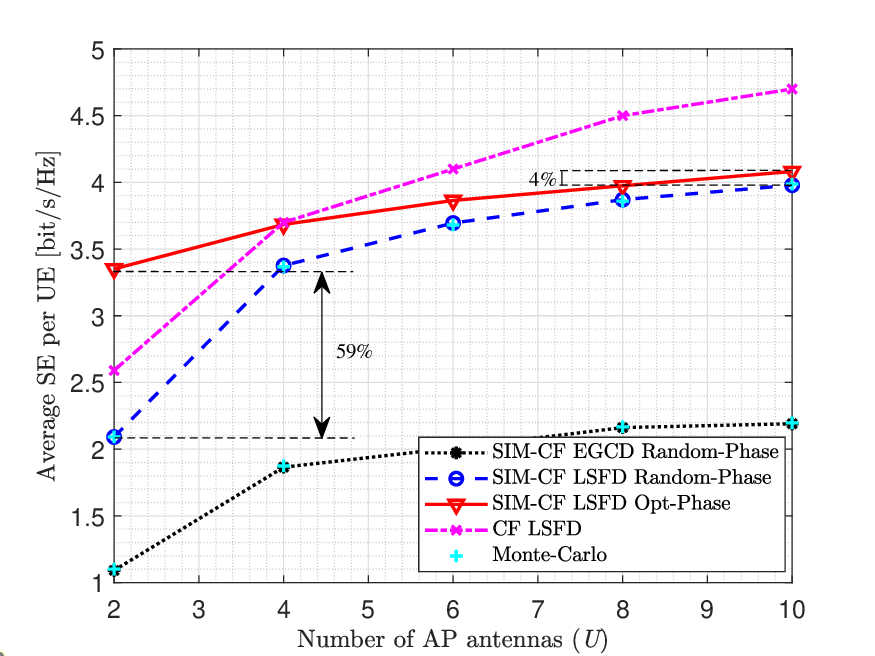}\vspace{-0.2cm}
\caption{Average SE per UE against different numbers of AP antennas over EGCD/LSFD with AP full power transmission ($L = 10$, $M = 5$, $N = 25$, $K = 5$, $\tau_p = 4$, ${d_x} = {d_y} = {1 \mathord{\left/
 {\vphantom {1 2}} \right.
 \kern-\nulldelimiterspace} 2}\lambda$).}\label{Fig_U}\vspace{-0.4cm}
\end{figure}
Fig.~\ref{Fig_U} illustrates the average SE per UE against different numbers of AP antennas over EGCD/LSFD with AP full power transmission. It is evident that increasing the number of AP antennas leads to a system performance increase. Furthermore, when the number of AP antennas is $L = 2$ and $L = 10$, the performance of the SIM-enhanced CF mMIMO system with optimized phase shifts improves by 59\% and 4\%, respectively, compared to the performance with random phase shifts. In particular, a 2-antenna SIM-enhanced AP can achieve SE performance close to that of a traditional 4-antenna AP. Moreover, when the number of AP antennas is limited, the SE performance of the optimized phase shift SIM-enhanced CF mMIMO system is significantly higher than that of the CF mMIMO system. However, as the number of antennas increases, the performance of the CF mMIMO system gradually surpasses that of the SIM-enhanced CF mMIMO system. It also reveals the benefits of SIM to simplify transceiver configurations and enhance system performance when the number of transceiver antennas is limited \cite{an2023stacked}. By contrast, the advantages of SIM become less apparent when the number of transceiver antennas is sufficiently large.

\subsection{Impact of SIM Parameters}

\begin{figure}[t]
\centering
\includegraphics[scale=0.55]{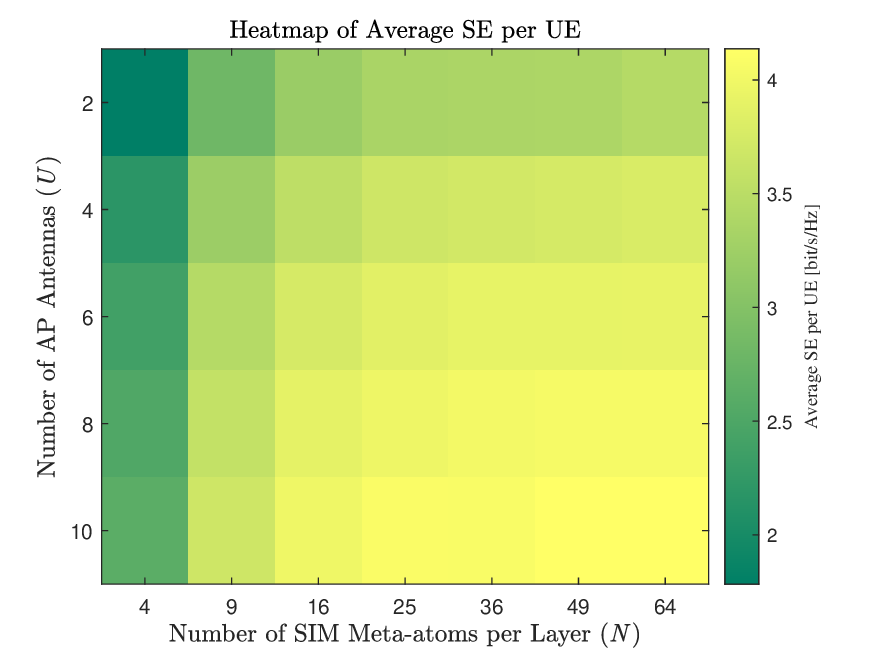}\vspace{-0.2cm}
\caption{Heatmap of average SE per UE against different numbers of AP antennas and SIM meta-atoms per layer over LSFD and wave-based beamforming optimization with AP full power transmission ($L = 10$, $M = 5$, $K = 5$, $\tau_p = 4$, ${d_x} = {d_y} = {1 \mathord{\left/
 {\vphantom {1 2}} \right.
 \kern-\nulldelimiterspace} 2}\lambda$).}\label{Fig_N_U_2D}\vspace{-0.4cm}
\end{figure}
Fig.~\ref{Fig_N_U_2D} illustrates the average SE per UE against different numbers of AP antennas and SIM meta-atoms per layer over LSFD and wave-based beamforming optimization with AP full power transmission. It is evident that increasing either the number of SIM meta-atoms or AP antennas can enhance system performance. Furthermore, the effect of increasing the number of SIM meta-atoms on system performance is more pronounced than that of increasing the number of AP antennas. However, when both the number of AP antennas and SIM meta-atoms are large, further increases in their numbers yield only marginal gains in system performance.

\begin{figure}[t]
\centering
\includegraphics[scale=0.55]{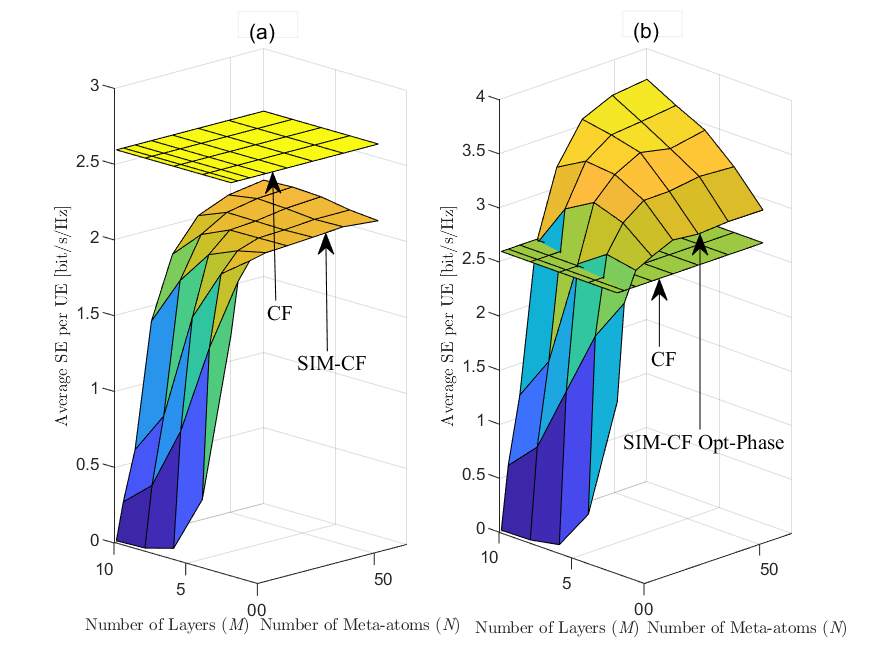}\vspace{-0.2cm}
\caption{3D figure of average SE per UE against different numbers of SIM layers and meta-atoms per layer over LSFD with AP full power transmission: (a) SIM-enhanced CF mMIMO system with random wave-based beamforming. (b) SIM-enhanced CF mMIMO system with wave-based beamforming optimization ($L = 10$, $K = 5$, $U = 2$, $\tau_p = 4$, ${d_x} = {d_y} = {1 \mathord{\left/
 {\vphantom {1 2}} \right.
 \kern-\nulldelimiterspace} 2}\lambda$).}\label{Fig_N_M_3D}\vspace{-0.4cm}
\end{figure}
Fig.~\ref{Fig_N_M_3D} illustrates the 3D figure of average SE per UE against different numbers of SIM layers and meta-atoms per layer over LSFD with AP full power transmission. It is clear from Fig.~\ref{Fig_N_M_3D} (a) that if random wave-based beamforming is adopted, even with LSFD, SIMs cannot bring performance gains to the CF mMIMO system and the performance will rapidly deteriorate with the increase in the number of SIM layers, especially when the number of meta-atoms per layer is small. However, as shown in Fig.~\ref{Fig_N_M_3D} (b), by optimizing SIM wave-based beamforming, the performance of the SIM-enhanced CF mMIMO system significantly surpasses that of the CF mMIMO system. Moreover, increasing the number of layers and meta-atoms in SIM can further enhance system performance.

\begin{figure}[t]
\centering
\includegraphics[scale=0.55]{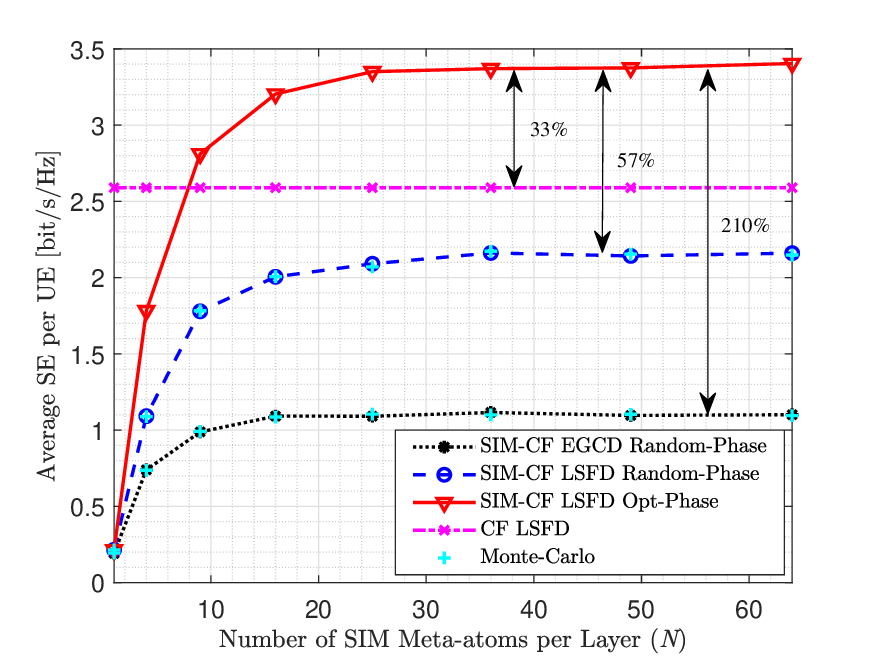}\vspace{-0.2cm}
\caption{Average SE per UE against different numbers of SIM meta-atoms per layer over EGCD/LSFD with AP full power transmission ($L = 10$, $M = 5$, $K = 5$, $U = 2$, $\tau_p = 4$, ${d_x} = {d_y} = {1 \mathord{\left/
 {\vphantom {1 2}} \right.
 \kern-\nulldelimiterspace} 2}\lambda$).}\label{Fig_N}\vspace{-0.4cm}
\end{figure}
Fig.~\ref{Fig_N} illustrates the average SE per UE against different numbers of SIM meta-atoms per layer over EGCD/LSFD with AP full power transmission. It is clear that increasing the number of SIM meta-atoms per layer is always beneficial in terms of the uplink SE. As the number of SIM meta-atoms $N>9$, the performance of LSFD with SIM wave-based beamforming optimization surpasses that of the CF mMIMO system. When performance stabilizes, the LSFD with phase-optimized SIM-enhanced CF mMIMO system performance improves by 33\%, 57\%, and 210\% compared to CF mMIMO, random phase shifts, and EGCD with random phase shifts, respectively. This indicates that a moderate increase in the number of meta-atoms significantly enhances system performance, while excessive meta-atoms may result in diminishing returns. Moreover, the computational complexity and system dimensions escalate exponentially. Hence, a strategy of continuously increasing the number of meta-atoms may not yield the most efficient system design, suggesting that an optimal number of meta-atoms exists for balancing performance gains against system complexity and resource usage.

\begin{figure}[t]
\centering
\includegraphics[scale=0.55]{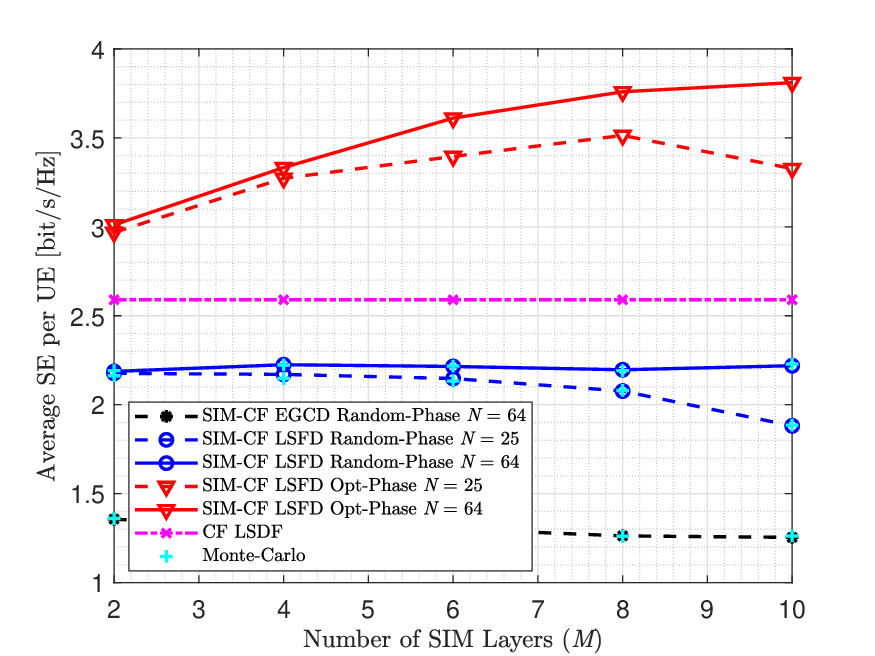}\vspace{-0.2cm}
\caption{Average SE per UE against different numbers of SIM layers and meta-atoms per layer over EGCD/LSFD with AP full power transmission ($L = 10$, $K = 5$, $U = 2$, $\tau_p = 4$,${d_x} = {d_y} = {1 \mathord{\left/
 {\vphantom {1 2}} \right.
 \kern-\nulldelimiterspace} 2}\lambda$).}\label{Fig_N_M}\vspace{-0.4cm}
\end{figure}

%\begin{figure}[t]
%\centering
%\includegraphics[scale=0.4]{Fig_R.eps}\vspace{-0.2cm}
%\caption{Average SE per UE against different spacing between adjacent transmit meta-atoms over LSFD with AP full power transmission ($L = 10$, $M = 5$, $N = 64$, $K = 5$, $U = 2$, $\tau_p = 4$).}\label{Fig_R}\vspace{-0.2cm}
%\end{figure}

Fig.~\ref{Fig_N_M} illustrates the average SE per UE against different numbers of SIM layers and meta-atoms per layer over EGCD/LSFD with AP full power transmission. It is clear that in the absence of SIM wave-based optimization, merely processing signals at the AP receiver leads to a deterioration in system performance, particularly when the number of meta-atoms per SIM layer is minimal. This degradation becomes more pronounced with an increase in the number of SIM layers, indicating an interaction between layer density and meta-atom count per layer. Conversely, employing the proposed wave-based optimization method markedly improves system performance. Specifically, when the number of meta-atoms per layer is $N=25$, optimal system performance is achieved with $M=8$ SIM layers. Beyond this point, additional layers result in a degradation in performance, delineating a distinct peak in the system's efficiency curve. However, an increase in the number of meta-atoms to $N=64$ allows the system performance to continue improving up to $M=10$ layers. This observation underscores the importance of maintaining a sufficient number of SIM meta-atoms per layer; insufficient meta-atoms per layer limit the system's capacity to benefit from additional layers. Consequently, increasing the number of meta-atoms per layer not only enhances the optimal number of effective SIM layers but also significantly boosts overall system performance.

\renewcommand\arraystretch{1.5}
\begin{table}[t]
\caption{Average SE against Different Spacing between Adjacent Meta-atoms over LSFD with AP Full Power Transmission. \label{Table_Fronthual}}
\centering
\begin{tabular}{!{\vrule width1.2pt}  m{3.3 cm}<{\centering} !{\vrule width1pt}  m{0.8 cm}<{\centering} !{\vrule width1pt}  m{0.8 cm}<{\centering} !{\vrule width1pt}  m{0.8 cm}<{\centering} !{\vrule width1pt}  m{0.8 cm}<{\centering} !{\vrule width1.2pt}}
   \Xhline{1.2pt}
   \rowcolor{gray!50} \bf \makecell[c]{Space between\\ Meta-atoms}& \bf \makecell[c]{$\lambda$}& \bf ${\lambda  \mathord{\left/
 {\vphantom {\lambda  2}} \right.
 \kern-\nulldelimiterspace} 2}$\(\bigstar\)& \bf \makecell[c]{${\lambda  \mathord{\left/
 {\vphantom {\lambda  4}} \right.
 \kern-\nulldelimiterspace} 4}$} & \bf \makecell[c]{${\lambda  \mathord{\left/
 {\vphantom {\lambda  8}} \right.
 \kern-\nulldelimiterspace} 8}$}\\
\hline
\bf \makecell[c]{SE opt-phase\\ (bit/s/Hz)}&3.06&\bf3.43&3.37&3.32 \\  
\cline{1-5}  
\bf \makecell[c]{Performance gap\\ w.r.t to ${\lambda  \mathord{\left/
 {\vphantom {\lambda  2}} \right.
 \kern-\nulldelimiterspace} 2}$ SE} &-11\%&-&-2\%&-3\% \\
\cline{1-5}
\bf \makecell[c]{SE random-phase\\ (bit/s/Hz)}&1.85&\bf2.15&1.99&1.79\\
\cline{1-5}
\bf \makecell[c]{Performance gap\\ w.r.t to ${\lambda  \mathord{\left/
 {\vphantom {\lambda  2}} \right.
 \kern-\nulldelimiterspace} 2}$ SE}&-14\%&-&-7\%&-17\%\\
 \Xhline{1.2pt}
\end{tabular}
\end{table}
Table~\ref{Table_Fronthual} illustrates the average SE per UE against different spacing between adjacent transmit meta-atoms over LSFD with AP full power transmission. We use $d$ to denote the space between adjacent meta-atoms. The results indicate that both large or small spacings between adjacent meta-atoms lead to a decrease in system performance, with optimal performance achieved at half-wavelength. Also, with random and optimized phase shifts, the performance at ${\lambda  \mathord{\left/
 {\vphantom {\lambda  8}} \right.
 \kern-\nulldelimiterspace} 8}$ decreases by 3\% and 17\% respectively compared to ${\lambda  \mathord{\left/
 {\vphantom {\lambda  2}} \right.
 \kern-\nulldelimiterspace} 2}$. This reveals that both large and small element spacings can lead to similarity between transmission coefficients, resulting in undesired channel correlation, which in turn leads to poor channel matching. However, optimizing SIM phase shifts can mitigate the performance degradation caused by spatial correlation.

\subsection{Impact of Power Control Schemes}

\begin{figure}[t]
\centering
\includegraphics[scale=0.55]{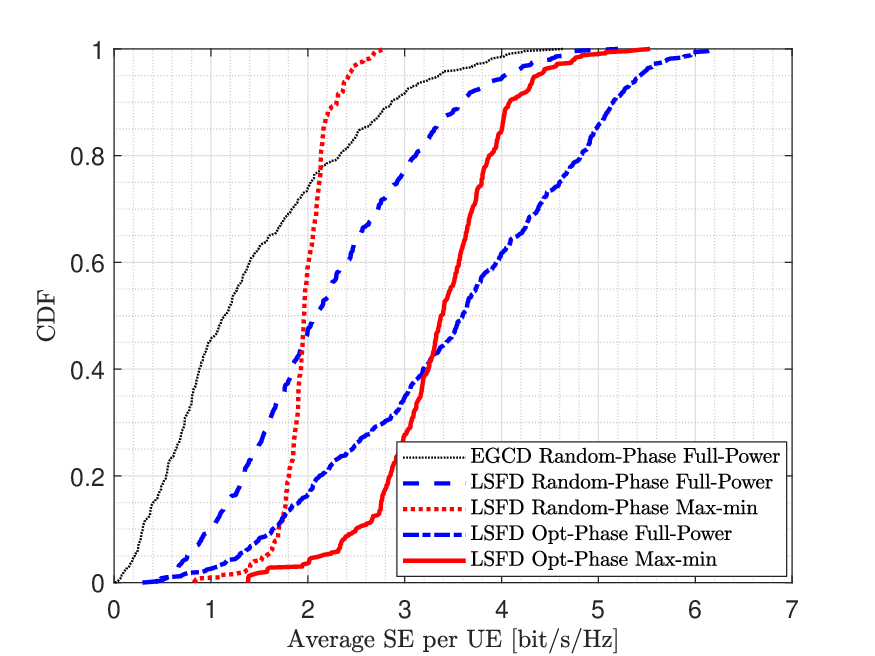}\vspace{-0.2cm}
\caption{CDF of the average SE per UE for the SIM-enhanced CF mMIMO system against different power control methods over EGCD/LSFD ($L = 10$, $M = 10$, $N = 64$, $K = 5$, $U = 2$, $\tau_p = 4$, ${d_x} = {d_y} = {1 \mathord{\left/
 {\vphantom {1 2}} \right.
 \kern-\nulldelimiterspace} 2}\lambda$).}\label{Fig_PC_CDF2}\vspace{-0.4cm}
\end{figure}
Fig.~\ref{Fig_PC_CDF2} illustrates the CDF of the average SE per UE for the SIM-enhanced CF mMIMO system against different power control methods over EGCD/LSFD. It is clear that compared to full-power transmission with random phase shifts under LSFD, the full-power transmission with opt-phase under LSFD, max-min power control with random phase shifts under LSFD, and max-min power control with opt-phase under LSFD, the performance of the 95\%-likely per-user SE improved by 70\%, 106\%, and 188\%, respectively. The results demonstrate that the proposed max-min SE power control scheme can significantly enhance the performance of UEs with poor conditions. Additionally, it is clearly observable that phase shift optimization has a more pronounced effect on improving the performance of better-performing UEs while the max-min SE power control scheme sacrifices the performance of better users. Therefore, in practice, it is necessary to choose different power control strategies based on actual UE requirements.

\begin{figure}[t]
\centering
\includegraphics[scale=0.55]{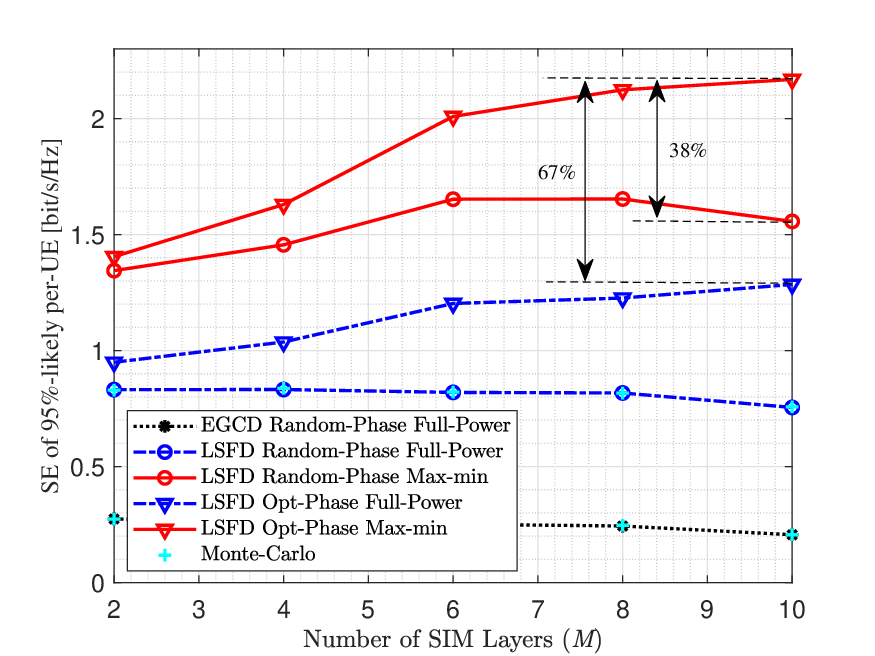}\vspace{-0.2cm}
\caption{Average SE per UE for the SIM-enhanced CF mMIMO system against different numbers of SIM layers and various power control methods over EGCD/LSFD ($L = 10$, $N = 64$, $K = 5$, $U = 2$, $\tau_p = 4$, ${d_x} = {d_y} = {1 \mathord{\left/
 {\vphantom {1 2}} \right.
 \kern-\nulldelimiterspace} 2}\lambda$).}\label{Fig_PC_M2}\vspace{-0.4cm}
\end{figure}

\begin{figure}[t]
\centering
\includegraphics[scale=0.55]{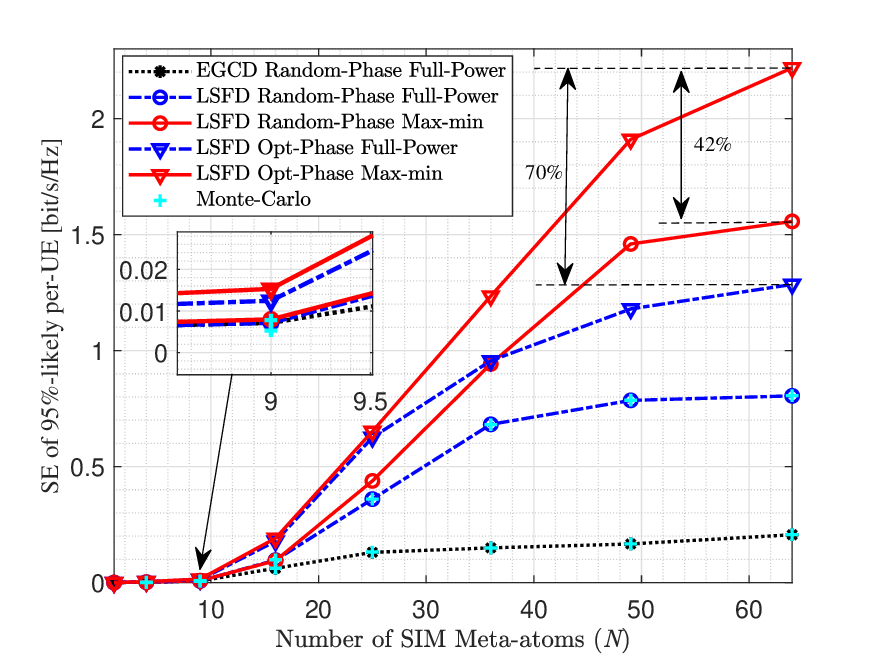}\vspace{-0.2cm}
\caption{Average SE per UE for the SIM-enhanced CF mMIMO system against different numbers of SIM meta-atoms per layer and various power control methods over EGCD/LSFD ($L = 10$, $M = 10$, $K = 5$, $U = 2$, $\tau_p = 4$, ${d_x} = {d_y} = {1 \mathord{\left/
 {\vphantom {1 2}} \right.
 \kern-\nulldelimiterspace} 2}\lambda$).}\label{Fig_PC_N2}\vspace{-0.4cm}
\end{figure}

Fig.~\ref{Fig_PC_M2} illustrates the average SE per UE for the SIM-enhanced CF mMIMO system against different numbers of SIM layers and various power control methods over EGCD/LSFD. It is clear that adopting phase shift optimization with power control can significantly enhance the performance of UEs with poor conditions, and this improvement becomes more pronounced with an increase in the number of SIM layers. Specifically, when SIM layers $M = 10$ and adopting max-min power control scheme and phase shift optimization, the performance of 95\%-likely per-UE SE improves 67\% and 38\% compared with only phase shift optimization and only max-min power control, respectively. In contrast, employing random phase shift settings along with full power transmission results in a degradation of performance for these UEs as the number of SIM layers increases. This deterioration occurs because random phases do not effectively exploit the beamforming potential of SIM meta-atoms, and the increased number of layers introduces additional signal attenuation.
Furthermore, while the use of random phase shift in conjunction with max-min power control initially leads to performance gains with an increasing number of SIM layers, this improvement plateaus and begins to decline beyond layers $M=6$. This decline is attributed to the inefficient use of beamforming capabilities by random phase shifts, coupled with attenuation losses exacerbated by the multilayer structure.

Fig.~\ref{Fig_PC_N2} illustrates the average SE per UE for the SIM-enhanced CF mMIMO system against different numbers of SIM meta-atoms per layer and various power control methods over EGCD/LSFD. It is clear that increasing the number of SIM meta-atoms results in a significant improvement in performance for the 95\%-likely per-user SE. However, when the number of meta-atoms $N<9$, the SE of poor UEs is nearly zero, and the advantages of the max-min power control scheme begin to manifest when the number of SIM meta-atoms $N>25$. 
When the number of SIM meta-atoms $N=64$, the SE performance with both phase and power optimization improves by 42\% and 70\%, respectively, compared to optimizing power only and optimizing phase only. 
Interestingly, when the meta-atoms $N<36$, phase-only optimization performs better than power-only optimization; the reverse is true when the meta-atoms $N>36$. 
This is due to the enhanced spatial diversity afforded by a greater number of SIM meta-atoms, which can be fully leveraged to improve UE performance even in the absence of phase shift optimization. Conversely, when the meta-atom count is low, the limited spatial diversity fails to effectively enhance UE performance, thus underscoring the critical role of power control.
This insight suggests that in practice when computational resources are limited, the number of SIM meta-atoms can guide the choice between phase optimization and power control design to ensure UE fairness.

\section{Conclusion}
In this paper, we introduce a novel SIM-enhanced CF mMIMO framework and investigate the performance of the considered system. First, closed-form SE expressions are derived using a two-layer architecture for signal processing, where MR combining is adopted by the APs, and LSFD/EGCD is adopted by the CPU. Second, an interference-based greedy algorithm for pilot allocation is introduced to mitigate inter-user interference. Third, based on the derived SE closed-form expression, an iterative optimization algorithm is proposed for designing the wave-based beamforming of SIM on a long-term scale to unlock its potential. This algorithm requires only the statistical CSI to configure SIMs once in each coherence block, thereby reducing the frequency of optimization needed for SIMs. Finally, a max-min SE power control algorithm is proposed to enhance the performance of the UEs with poor channel conditions. 
The results indicate that the proposed SIM wave-based beamforming can enhance the performance of SIM-enhanced CF mMIMO systems by 57\%, significantly surpassing traditional CF mMIMO systems. Additionally, the proposed power control algorithm improves the SE performance of 95\%-likely per-user by 106\% compared to full-power transmission. 
In particular, with the introduction of SIM, deploying only 10 APs in the SIM-enhanced CF mMIMO system achieves SE performance that surpasses that of a traditional CF mMIMO system with 15 APs. 
Also, a 2-antenna SIM-enhanced AP can achieve SE performance close to that of a traditional 4-antenna AP, thereby reducing antenna costs.
Besides, the size of the SIM meta-atoms also affects system performance with optimal performance gains observed at half-wavelength spacing. Finally, deploying SIMs equipped with a large number of meta-atoms can serve as an alternative to APs and AP antennas, enhancing the performance of the CF mMIMO system.

\begin{appendices}
\section{Proof of Theorem 1}
The expectations in \eqref{SINR_k_close} are evaluated in this section. We start by considering the numerator term as follows
\begin{align}
&\mathbb{E}\left\{ {\sum\limits_{l = 1}^L {a_{l,k}^ * } {\bf{v}}_{l,k}^{\rm{H}}{\bf{W}}_{l,1}^{\rm{H}}{\bf{G}}_l^{\rm{H}}{\bf{h}}_{{\rm{SI}}{{\rm{M}}_l},k}^{}} \right\} \notag\\
&= {\bf{a}}_k^{\rm{H}}\mathbb{E}\left\{ {{{\left[\ldots, {{\bf{\hat h}}_{l,k}^{\rm{H}}{\bf{W}}_{l,1}^{\rm{H}}{\bf{G}}_l^{\rm{H}}{\bf{h}}_{{\rm{SI}}{{\rm{M}}_l},j}^{}} ,\ldots\right]}^{\rm{T}}}} \right\}.
\end{align}
Then, for $\mathbb{E}\left\{ {{\bf{\hat h}}_{l,k}^{\rm{H}}{\bf{W}}_{l,1}^{\rm{H}}{\bf{G}}_l^{\rm{H}}{\bf{h}}_{{\rm{SI}}{{\rm{M}}_l},j}^{}} \right\}$, we can obtain that 
\begin{align}
&\mathbb{E}\left\{ {{\bf{\hat h}}_{l,k}^{\rm{H}}{\bf{W}}_{l,1}^{\rm{H}}{\bf{G}}_l^{\rm{H}}{\bf{h}}_{{\rm{SI}}{{\rm{M}}_l},j}^{}} \right\} \notag\\
&= \mathbb{E}\left\{ {\left( {{\bf{W}}_{l,1}^{\rm{H}}{\bf{G}}_l^{\rm{H}}{\bf{\bar h}}_{{\rm{SI}}{{\rm{M}}_l},k}^{}{e^{j{\varphi _{{\rm{SI}}{{\rm{M}}_l},k}}}} + } \right.} \right. \notag\\
&\left. {{{\left. {\sqrt {{{\hat p}_k}} {{\bf{R}}_{l,k}}{\bf{\Psi }}_{l,k}^{ - 1}\left( {{\bf{y}}_{l,k}^p - {\bf{\bar y}}_{l,k}^p} \right)} \right)}^{\rm{H}}}{\bf{W}}_{l,1}^{\rm{H}}{\bf{G}}_l^{\rm{H}}{\bf{h}}_{{\rm{SI}}{{\rm{M}}_l},j}^{}} \right\} \notag\\
&= {\left\| {{\bf{W}}_{l,1}^{\rm{H}}{\bf{G}}_l^{\rm{H}}{\bf{\bar h}}_{{\rm{SI}}{{\rm{M}}_l},k}^{}} \right\|^2} + {\rm{tr}}\left( {{p_k}{\tau _p}{\bf{R}}_{l,k}^{}{\bf{\Psi }}_{l,k}^{ - 1}{\bf{R}}_{l,k}^{}} \right).
\end{align}

For the denominator part. We first compute the noise term as
\begin{align}
&\mathbb{E}\left\{ {{{\left\| {\sum\limits_{l = 1}^L {a_{l,k}^ * } {\bf{v}}_{l,k}^{\rm{H}}} \right\|}^2}} \right\} \notag\\
&= {\bf{a}}_k^{\rm{H}}{\rm{diag}}\left( \mathbb{E}{\left\{ {{{\left\| {{\bf{v}}_{1,k}^{\rm{H}}} \right\|}^2}} \right\}, \ldots ,\mathbb{E}\left\{ {{{\left\| {{\bf{v}}_{L,k}^{\rm{H}}} \right\|}^2}} \right\}} \right) {\bf{a}}_k \notag\\
& = {\bf{a}}_k^{\rm{H}}\!{\rm{diag}}\left( {\cdots,{\rm{tr}}\left( {{\hat p_k}{\tau _p}{{\bf{\Omega }}_{l,k}}} \right) + {{\left\| {{{{\bf{\bar h}}}_{l,k}}} \right\|}^2}}, \cdots\!\right){{\bf{a}}_k}.
\end{align}
The interference term in the denominator of \eqref{SINR_k_close} is
\begin{align}
&\mathbb{E}\left\{ {{{\left| {\sum\limits_{l = 1}^L {a_{l,k}^{}{\bf{\hat h}}_{l,k}^{\rm{H}}{{\bf{h}}_{l,j}}} } \right|}^2}} \right\}\notag\\
&= \sum\limits_{l = 1}^L {\sum\limits_{l' = 1}^L {{a_{l,k}}a_{l',k}^ * } } \mathbb{E}\left\{ {{{\left( {{\bf{\hat h}}_{l,k}^{\rm{H}}{{\bf{h}}_{l,j}}} \right)}^{\rm{H}}}\left( {{\bf{\hat h}}_{l',k}^{\rm{H}}{{\bf{h}}_{l',j}}} \right)} \right\},
\end{align}
where $\mathbb{E}\left\{ {{{\left( {{\bf{\hat h}}_{l,k}^{\rm{H}}{{\bf{h}}_{l,j}}} \right)}^{\rm{H}}}\left( {{\bf{\hat h}}_{l',k}^{\rm{H}}{{\bf{h}}_{l',j}}} \right)} \right\}$ is computed for all possible APs and UEs combinations. We utilize the independence of channel estimation at different APs. When $l \ne l',j \notin {{\cal P}_k}$, we obtain $\mathbb{E}\left\{ {{{\left( {{\bf{\hat h}}_{l,k}^{\rm{H}}{{\bf{h}}_{l,j}}} \right)}^{\rm{H}}}\left( {{\bf{\hat h}}_{l',k}^{\rm{H}}{{\bf{h}}_{l',j}}} \right)} \right\} = 0$. For $l \ne l',j \in {{\cal P}_k}\backslash \left\{ k \right\}$, we derive $\mathbb{E}\left\{ {{{\left( {{\bf{\hat h}}_{l,k}^{\rm{H}}{{\bf{h}}_{l,j}}} \right)}^{\rm{H}}}\left( {{\bf{\hat h}}_{l',k}^{\rm{H}}{{\bf{h}}_{l',j}}} \right)} \right\} = \mathbb{E}\left\{ {{\bf{\hat h}}_{l,j}^{\rm{H}}{{{\bf{\hat h}}}_{l,k}}} \right\}\mathbb{E}\left\{ {{\bf{\hat h}}_{l',k}^{\rm{H}}{{{\bf{\hat h}}}_{l',j}}} \right\}$, where
\begin{align}
\mathbb{E}\left\{ {{\bf{\hat h}}_{l,j}^{\rm{H}}{{{\bf{\hat h}}}_{l,k}}} \right\} = \sqrt {{{\hat p}_k}{{\hat p}_j}} {\tau _p}{\rm{tr}}\left( {{\bf{R}}_{l,k}{\bf{\Psi }}_{l,k}^{ - 1}{\bf{R}}_{l,j}} \right),
\end{align}
since $\mathbb{E}\left\{ {{{{\bf{\bar h}}}^{{\rm{H}}}_{l,k}}{{{\bf{\bar h}}}_{l,j}}{e^{ - j{\varphi _{{{\rm{SIM}}_l},k}}}}{e^{j{\varphi _{{{\rm{SIM}}_l},j}}}}} \right\} = 0$ and $\mathbb{E}\left\{ {\sqrt {{\hat p_k}{\tau _p}} \left({\bf{R}}_{l,k}{\bf{\Psi }}_{l,k}^{ - 1}\left( {{\bf{y}}_{l,k}^p - {\bf{\bar y}}_{l,k}^p} \right)\right)^{\rm{H}}{{{\bf{\bar h}}}_{l,j}}{e^{j{\varphi _{{{\rm{SIM}}_l},j}}}}} \right\} = 0$. We repeat the same calculation for AP $l'$ and obtain
\begin{align}
&\mathbb{E}\left\{ {{{\left( {{\bf{\hat h}}_{l,k}^{\rm{H}}{{\bf{h}}_{l,j}}} \right)}^{\rm{H}}}\left( {{\bf{\hat h}}_{l',k}^{\rm{H}}{{\bf{h}}_{l',j}}} \right)} \right\} \notag\\
&= {{\hat p}_k}{{\hat p}_j}\tau_p^2{\rm{tr}}\left( {{\bf{R}}_{l,k}{\bf{\Psi }}_{l,k}^{ - 1}{\bf{R}}_{l,j}} \right){\rm{tr}}\left( {{\bf{R}}_{l',j}{\bf{\Psi }}_{l',k}^{ - 1}{\bf{R}}_{l',k}} \right).
\end{align}
For another case $l \ne l',j = k$, we obtain
\begin{align}
&\mathbb{E}\left\{ {{{\left( {{\bf{\hat h}}_{l,k}^{\rm{H}}{{\bf{h}}_{l,j}}} \right)}^{\rm{H}}}\left( {{\bf{\hat h}}_{l',k}^{\rm{H}}{{\bf{h}}_{l',j}}} \right)} \right\} \notag\\
&= \hat p_k^2\tau _p^2{\rm{tr}}\left( {{{\bf{\Omega }}_{l,k}}} \right){\rm{tr}}\left( {{{\bf{\Omega }}_{l',k}}} \right) + {\rm{tr}}\left( {{{{\bf{\bar h}}}_{l,k}}{\bf{\bar h}}_{l,k}^{\rm{H}}} \right){\rm{tr}}\left( {{{{\bf{\bar h}}}_{l',k}}{\bf{\bar h}}_{l',k}^{\rm{H}}} \right)\notag\\
&+ {{\hat p}_k}{\tau_p}{\rm{tr}}\left( {{{\bf{\Omega }}_{l',k}}} \right){\rm{tr}}\left( {{{{\bf{\bar h}}}_{l,k}}{\bf{\bar h}}_{l,k}^{\rm{H}}} \right) \notag\\
&+ {{\hat p}_k}{\tau _p}{\rm{tr}}\left( {{{\bf{\Omega }}_{l,k}}} \right){\rm{tr}}\left( {{{{\bf{\bar h}}}_{l',k}}{\bf{\bar h}}_{l',k}^{\rm{H}}} \right).
\end{align}
Similarly for $l = l',j = k$, we utilize the same method as in \cite{ozdogan2019performance} that yields
\begin{align}
&\mathbb{E}\left\{ {{{\left| {{\bf{\hat h}}_{l,k}^{\rm{H}}{{\bf{h}}_{l,k}}} \right|}^2}} \right\} = \hat p_k^2\tau _p^2{\left| {{\rm{tr}}\left( {{{\bf{\Omega }}_{l,k}}} \right)} \right|^2} \notag\\
&+ {{\hat p}_k}{\tau _p}{\rm{tr}}\left( {{{\bf{\Omega }}_{l,k}}{\bf{R}}_{l,k}} \right) + {\bf{\bar h}}_{l,k}^{\rm{H}}{\bf{R}}_{l,k}{{{\bf{\bar h}}}_{l,k}} + {{\hat p}_k}{\tau _p}{\bf{\bar h}}_{l,k}^{\rm{H}}{{\bf{\Omega }}_{l,k}}{{{\bf{\bar h}}}_{l,k}} \notag\\
&+ 2{{\hat p}_k}{\tau _p}{\rm{tr}}\left( {{{\bf{\Omega }}_{l,k}}} \right){\bf{\bar h}}_{l,k}^{\rm{H}}{{{\bf{\bar h}}}_{l,k}} + {\rm{tr}}{\left( {{{{\bf{\bar h}}}_{l,k}}{\bf{\bar h}}_{l,k}^{\rm{H}}} \right)^2}.
\end{align}
Then, for the case $l = l',j \notin {{\cal P}_k}$, we obtain
\begin{align}
&\mathbb{E}\left\{ {{{\left| {{\bf{\hat h}}_{l,k}^{\rm{H}}{{\bf{h}}_{l,j}}} \right|}^2}} \right\} = {\hat p_k}{\tau _p}{\rm{tr}}\left( {{\bf{R}}_{l,j}{{\bf{\Omega }}_{l,k}}} \right)\notag\\
&+ {\bf{\bar h}}_{l,k}^{\rm{H}}{\bf{R}}_{l,j}{{{\bf{\bar h}}}_{l,k}} + {\hat p_k}{\tau _p}{\bf{\bar h}}_{l,j}^{\rm{H}}{{\bf{\Omega }}_{l,k}}{{{\bf{\bar h}}}_{l,j}} + {\left| {{\bf{\bar h}}_{l,k}^{\rm{H}}{{{\bf{\bar h}}}_{l,j}}} \right|^2}.
\end{align}
For $l = l',j \in {{\cal P}_k}\backslash \left\{ k \right\}$, we obtain
\begin{align}
&\mathbb{E}\left\{ {{{\left| {{\bf{\hat h}}_{l,k}^{\rm{H}}{{\bf{h}}_{l,j}}} \right|}^2}} \right\} = {{\hat p}_k}{\tau _p}{\rm{tr}}\left( {{{\bf{\Omega }}_{l,k}}{\bf{R}}_{l,j}} \right){\rm{ + tr}}\left( {{{{\bf{\bar h}}}_{l,k}}{\bf{\bar h}}_{l,k}^{\rm{H}}{\bf{R}}_{l,j}} \right)\notag\\
&+ {\left| {{\bf{\bar o}}_{l,k}^{\rm{H}}{{{\bf{\bar h}}}_{l,j}}} \right|^2} + {{\hat p}_k}{{\hat p}_j}\tau _p^2{\left| {{\rm{tr}}\left( {{\bf{R}}_{l,k}{\bf{\Psi }}_{l,k}^{ - 1}{\bf{R}}_{l,k}} \right)} \right|^2}.
\end{align}
Finally, we can derive the expectation of \eqref{SINR_k_close}.

\end{appendices}

%\vspace{-0.25cm}
\bibliographystyle{IEEEtran}
%\newpage
%\vspace{-0.1cm}
\bibliography{IEEEabrv,Ref}

\end{document}